\documentclass[iop]{emulateapj}
\usepackage{apjfonts}
\usepackage{natbib}

\newcommand{\chandra}{{\it Chandra\/}}
\newcommand{\nustar}{{\it NuSTAR\/}}
\newcommand{\flux}{{erg~cm$^{-2}$~s$^{-1}$}}
\newcommand{\lum}{{erg~s$^{-1}$}}

\usepackage{amsmath}

\begin{document}
\title{Weak Hard \hbox{X-ray} Emission from Two Broad Absorption Line
Quasars Observed with \nustar: Compton-thick Absorption or Intrinsic
X-ray Weakness?}
\author{
B.~Luo,\altaffilmark{1,2}
W.~N.~Brandt,\altaffilmark{1,2}
D.~M.~Alexander,\altaffilmark{3}
F.~A.~Harrison,\altaffilmark{4}
D.~Stern,\altaffilmark{5}
F.~E.~Bauer,\altaffilmark{6,7}
S.~E.~Boggs,\altaffilmark{8}
F.~E.~Christensen,\altaffilmark{9}
A.~Comastri,\altaffilmark{10}
W.~W.~Craig,\altaffilmark{11,8}
A.~C.~Fabian,\altaffilmark{12}
D.~Farrah,\altaffilmark{13}
F.~Fiore,\altaffilmark{14}
F.~Fuerst,\altaffilmark{4}
B.~W.~Grefenstette,\altaffilmark{4}
C.~J.~Hailey,\altaffilmark{15}
R.~Hickox,\altaffilmark{3,16}
K.~K.~Madsen,\altaffilmark{4}
G.~Matt,\altaffilmark{17}
P.~Ogle,\altaffilmark{18}
G.~Risaliti,\altaffilmark{19,20}
C.~Saez,\altaffilmark{6}
S.~H.~Teng,\altaffilmark{21}
D.~J.~Walton,\altaffilmark{4}
\& W.~W.~Zhang\altaffilmark{22}}

\altaffiltext{1}{Department of Astronomy \& Astrophysics, 525 Davey Lab,
The Pennsylvania State University, University Park, PA 16802, USA}
\altaffiltext{2}{Institute for Gravitation and the Cosmos, The Pennsylvania State University, University Park, PA 16802, USA}
\altaffiltext{3}{Department of Physics, Durham University, South Road,
Durham DH1 3LE, UK}
\altaffiltext{4}{Cahill Center for Astronomy and Astrophysics, 
California Institute
of Technology, Pasadena, CA 91125, USA}
\altaffiltext{5}{Jet Propulsion Laboratory, California Institute of Technology,
Pasadena, CA 91109, USA}
\altaffiltext{6}{Pontificia Universidad Cat\'{o}lica de Chile,
Departamento de Astronom\'{\i}a y Astrof\'{\i}sica, Casilla 306,
Santiago 22, Chile}
\altaffiltext{7}{Space Science Institute, 4750 Walnut Street, Suite
205, Boulder, CO 80301, USA}
\altaffiltext{8}{Space Sciences Laboratory, University of California, Berkeley,
CA 94720, USA}
\altaffiltext{9}{DTU Space - National Space Institute, Technical University of
Denmark, Elektrovej 327, 2800 Lyngby, Denmark}
\altaffiltext{10}{INAF---Osservatorio Astronomico di Bologna, Via Ranzani 1,
Bologna, Italy}
\altaffiltext{11}{Lawrence Livermore National Laboratory, Livermore, CA
94550, USA}
\altaffiltext{12}{Institute of Astronomy, Madingley Road, Cambridge,
CB3 0HA, UK}
\altaffiltext{13}{Department of Physics, Virginia Tech, Blacksburg, VA 24061, USA}
\altaffiltext{14}{Osservatorio Astronomico di Roma, via Frascati 33, 00040 Monteporzio Catone, Italy}
\altaffiltext{15}{Columbia Astrophysics Laboratory, Columbia University, New
York, NY 10027, USA}
\altaffiltext{16}{Department of Physics and Astronomy, Dartmouth College, 6127 Wilder Laboratory, Hanover, NH 03755, USA}
\altaffiltext{17}{Dipartimento di Matematica e Fisica, Universit\`{a} degli Studi Roma Tre, via della Vasca Navale 84, 00146 Roma, Italy}
\altaffiltext{18}{IPAC, California Institute of Technology, Mail Code 220-6, Pasadena, CA 91125, USA}
\altaffiltext{19}{INAF -- Osservatorio Astrofisico di Arcetri, Largo E. Fermi 5, I-50125 Firenze, Italy}
\altaffiltext{20}{Harvard-Smithsonian Center for Astrophysics, 60 Garden St., Cambridge, MA 02138, USA}
\altaffiltext{21}{Observational Cosmology Laboratory, NASA Goddard Space Flight Center, Greenbelt, MD 20771, USA}
\altaffiltext{22}{NASA Goddard Space Flight Center, Greenbelt, MD 20771, USA}

\begin{abstract}
We present \nustar\ hard \hbox{X-ray} observations of 
two \hbox{X-ray} weak 
broad absorption line (BAL) quasars, PG~1004+130 (radio loud) 
and PG~1700+518 (radio quiet).
Many BAL quasars appear X-ray weak, probably due to absorption by the shielding
gas between the nucleus and the accretion-disk wind.
The two targets are among the optically brightest
BAL quasars, yet they are known to be significantly X-ray weak
at rest-frame 2--10~keV (\hbox{16--120} times fainter than typical quasars).
We would expect to obtain $\approx400$--600 hard X-ray ($\ga10$~keV) 
photons with \nustar, provided that these photons are not significantly
absorbed ($N_{\rm H}\la10^{24}$~cm$^{-2}$).
However, both BAL quasars are only detected in the softer 
\nustar\ bands (e.g., 4--20~keV) 
but not in its harder bands (e.g., 20--30~keV), suggesting
that either the shielding gas is highly Compton-thick 
or the two targets are intrinsically X-ray weak.
We constrain the column densities for both to be 
$N_{\rm H}\approx7\times10^{24}$~cm$^{-2}$
if the weak hard \hbox{X-ray} emission is caused by obscuration from
the shielding gas.
We discuss a few possibilities for how PG~1004+130 could have Compton-thick
shielding gas without strong Fe~K$\alpha$ line emission;
dilution from jet-linked X-ray emission is one likely explanation.
We also discuss the intrinsic \hbox{X-ray} weakness scenario based on a 
coronal-quenching model relevant to the shielding gas and disk wind
of BAL quasars. 
Motivated by our \nustar\ results, we 
perform a \chandra\ stacking analysis with the Large Bright
Quasar Survey 
BAL quasar sample and 
place statistical constraints upon the fraction of
intrinsically \hbox{X-ray} weak BAL quasars; this fraction is likely 
$17\textrm{--}40\%$.

\end{abstract}

\section{INTRODUCTION}

\subsection{Quasar Outflows and the \hbox{X-ray} Properties of Broad Absorption Line Quasars}
Fast outflows are a common feature of active galactic nuclei 
(AGNs) over more than four orders of magnitude in luminosity 
\citep[e.g.,][]{Reynolds1997,Crenshaw1999,Laor2002,Ganguly2008,Gibson2009}. 
AGN outflows appear to be a substantial component of the
nuclear environment, and their ubiquity suggests that
mass ejection is probably linked to or even required for mass accretion onto
a supermassive black hole (SMBH). For example, outflows could provide
a mechanism for expelling angular momentum from the accreting
material \citep[e.g.,][]{Emmering1992,Konigl1994}. 
Moreover, outflows in luminous AGNs may play an important
role in the feedback of SMBHs into typical massive galaxies
\citep[e.g.,][]{Dimatteo2005,Chartas2009,Sturm2011,Borguet2012,Rupke2013}. The outflowing material could
drive away sufficient gas from the host galaxy to quench both star
formation and SMBH growth, leading to the observed relations between
the mass of the SMBH and the properties of the galaxy bulge \citep[e.g.,][]{Gultekin2009}.

The strongest observational signature of outflows from luminous
AGNs (i.e., quasars) is
broad absorption lines (BALs; \citealt{Lynds1967})
in the ultraviolet (UV); these are seen
in \hbox{$\approx15\%$} of optically selected quasars 
\citep[e.g.,][]{Hewett2003,Trump2006,Gibson2009,Allen2011}. 
Aside from dust reddening, BAL quasars generally have indistinguishable continuum spectral
energy distributions (SEDs) from non-BAL quasars from the 
infrared (IR) to the UV
\citep[e.g.,][]{Gallagher2007,Lazarova2012}.
It has been
suggested that all/most quasars have BAL winds, with BALs being observed
only when 
inclination angles are large and the line of sight passes through 
the outflowing wind
(e.g., \citealt{Weymann1991,Ogle1999,DiPompeo2012}). 
The intrinsic fraction of BAL quasars, after correcting for
selection effects, is $\approx20\%$ 
(see, e.g., \citealt{Gibson2009} and references therein), suggesting
that the wind has an average covering factor of $\approx0.2$.\footnote{There
is likely a range of covering factors of BAL winds, and $\approx0.2$ is the 
average value. Quasars with winds having larger covering factors would have 
larger chances of being observed as BAL quasars.}
An alternative hypothesis is that BAL quasars represent an early
evolutionary stage of quasars and/or the appearance as a BAL quasar
might be related to the duty cycle of SMBH growth
\citep[e.g.,][]{Becker2000,Gregg2006}.

\begin{figure}
\centerline{
\includegraphics[scale=0.35]{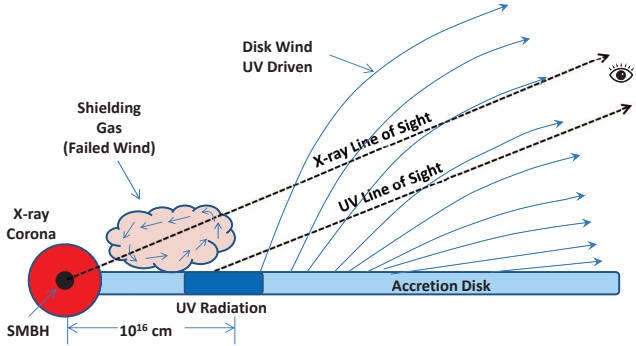}
}
\caption{A schematic diagram of the disk-wind scenario for BAL quasar outflows.
The wind is launched from the accretion disk at 
$\approx10^{16}$--10$^{17}$~cm and
is driven by UV radiation from the disk. 
BALs are observed when the line of sight passes through
the outflowing wind.
The shielding gas located 
at the base of the wind blocks the EUV and X-ray radiation from
the nucleus and prevents the wind from being overionized.
One origin for the shielding gas could be a ``failed wind'';
the small arrows in this gas represent the velocity field 
\citep[e.g.,][]{Proga2004}. 
For a standard
accretion disk, the disk region that emits most strongly in the UV 
has a radius $\la10^{16}$~cm; see Section~\ref{sec-physicalimp} below.
\label{fig-art}}
\end{figure}

A promising scenario for BAL quasar outflows is the accretion-disk wind model,
where the wind is launched from the disk at $\approx10^{16}$--10$^{17}$~cm and
is radiatively driven by UV line pressure \citep[e.g.,][]{Murray1995,Proga2000}.
Figure~\ref{fig-art} is a schematic illustration of the model.
UV absorption-line profiles predicted by this model are consistent with
observations of BAL quasars \citep[e.g.,][]{Proga2004}. 
Since line-driving becomes less efficient when the ionization state 
of the gas is too high, the disk-wind model has invoked 
``shielding'' material 
to prevent
the wind from being overionized by the extreme UV (EUV) and soft \hbox{X-ray} 
radiation from the innermost accretion disk and its corona. 
One proposed origin for the shielding gas 
is a ``failed wind'', which is located at the base of the UV-absorbing
wind
and consists of material that does not reach escape velocity due to 
overionization \citep[e.g.,][]{Proga2004,Sim2010}. The detailed geometry
of the shielding gas is still uncertain (e.g., it could also
perhaps ``hug'' the UV-absorbing wind).

Under the above scenario, AGNs with observable disk winds would appear 
X-ray weak due to absorption by the shielding gas. 
Indeed, BAL quasars are in general \hbox{X-ray} 
weak, and \hbox{X-ray} absorption is often
seen \citep[e.g.,][]{Gallagher2002a,Gallagher2006,Gibson2009}.
Moreover, the observed \hbox{X-ray} weakness
of BAL quasars is correlated with the absorption 
strength and maximum velocity of the
UV \ion{C}{4} BALs \citep[e.g.,][]{Gallagher2006,Gibson2009,Wu2010}, 
suggesting that the shielding gas 
does play a role in shaping the properties of the UV wind.
The \hbox{X-ray} absorption in BAL quasars often appears complex               
(i.e., not just a simple photoelectric absorption cutoff but
an ionized or partially covering absorber);
the measured
absorption column densities ($N_{\rm H}$) are typically in the range of
$10^{22}$ to $5\times10^{23}$~cm$^{-2}$, although absorption constraints
usually have significant
uncertainties due to limited photon statistics and poorly constrained 
absorption complexity
\citep[e.g.,][]{Gallagher2001,Gallagher2002a,Fan2009}.
X-ray absorption variability has been observed in a few BAL quasars 
on timescales of years
\citep[e.g.,][]{Gallagher2004,Miller2006,Chartas2009,Saez2012}, indicating 
that the shielding gas can be a dynamical structure
subject to rotational and outflow/inflow motions,
but the details of the dynamics remain unclear.\footnote{Substantial
BAL variability is also seen in BAL quasars on multi-year timescales
\citep[e.g.,][]{Gibson2008b,Gibson2010,Capellupo2011,Capellupo2012,Filiz2012},
but unfortunately correlations between \hbox{X-ray} and UV absorption variability have not been systematically explored.} Significant flux
variability has also been seen in a few objects \citep[e.g.,][]{Gallagher2004}.
However, when compared with the quasar population in general, 
BAL quasars do not show exceptional \hbox{X-ray} variability 
\citep{Saez2012}.
Overall, the nature of the shielding gas in BAL quasars remains poorly
constrained.

With the current \hbox{X-ray} spectra of BAL quasars observed by {\it ASCA}, 
\chandra, or
{\it XMM-Newton}, the \hbox{X-ray} absorption column density can be constrained
up to $\approx5\times10^{23}$~cm$^{-2}$. At higher column densities,
even photons at energies of \hbox{$\approx7$--10}~keV are severely absorbed, 
and
the observed \hbox{X-ray} spectra may be dominated by a scattered/reflected 
component.
It is thus difficult to determine the absorption properties for
heavily obscured ($N_{\rm H}\ga5\times10^{23}$~cm$^{-2}$)
or Compton-thick ($N_{\rm H}>1.5\times10^{24}$~cm$^{-2}$; see, e.g., \citealt{Comastri2004} for a review) objects. 
Supplemented with multiwavelength data, the level of \hbox{X-ray} weakness
can be estimated by comparing the SEDs 
of BAL quasars to those of typical quasars 
\citep[e.g.,][]{Gallagher2001,Gallagher2007,Miller2006}. Specifically, 
in such comparisons
the 
\hbox{X-ray-to-optical} power-law slope parameter ($\alpha_{\rm OX}$) is 
often used
\citep[e.g.,][]{Gallagher2002a,Gallagher2006}, 
which is a measure of the soft \hbox{X-ray} (2~keV) luminosity of a quasar
relative to its optical/UV luminosity. This parameter is known to be correlated
with the 2500~\AA\ monochromatic luminosity 
($L_{\rm 2500~{\textup{\AA}}}$) of the quasar 
\citep[e.g.,][]{Steffen2006}, and thus $\Delta\alpha_{\rm OX}$,
defined as the difference
between the observed $\alpha_{\rm OX}$ and the one expected from 
$L_{\rm 2500~{\textup{\AA}}}$ ($\Delta\alpha_{\rm OX}=\alpha_{\rm OX,obs}-\alpha_{\rm OX,exp}$), indicates the level \hbox{X-ray} weakness of the source.

For many BAL quasars, 
the observed \hbox{X-ray} weakness can be attributed entirely to absorption.
It is clear that
their \hbox{$\ga5$~keV} photons have penetrated the obscuring material,
and, after correcting for the moderate absorption,
their \hbox{X-ray} fluxes recover to nominal levels 
(\hbox{$\Delta\alpha_{\rm OX,corr}\approx0$}).
In these cases,
the absorption-corrected 2~keV flux, and thus 
$\alpha_{\rm OX,corr}$ and $\Delta\alpha_{\rm OX,corr}$, are generally estimated
assuming a power-law spectrum (e.g., with a nominal
photon index $\Gamma=1.8$) and normalizing it to the hard \hbox{X-ray} 
(e.g., rest-frame $\ga5$~keV) continuum 
\citep[e.g.,][]{Gallagher2006,Fan2009}.
For some of the sources with better spectra, absorption corrections 
can be derived directly via spectral fitting \citep[e.g.,][]{Gallagher2002a,Grupe2003,Shemmer2005,Giustini2008}.
It has also been found that the effective photon index determined
from hardness ratios is correlated with the level of 
X-ray weakness ($\Delta\alpha_{\rm OX}$), i.e., 
X-ray weaker sources are harder, suggesting that 
absorption plays an important role in causing the X-ray weakness
of the general BAL quasar population \citep[e.g.,][]{Gallagher2006}.

On the other hand, some BAL quasars show significant
\hbox{X-ray} weakness which cannot be accounted for by the {\it apparent\/} X-ray
absorption determined using $<10$~keV data. 
In these cases the \hbox{X-ray} continuum levels are still factors
of \hbox{$\approx5$--20} lower than the expected SED levels after 
absorption corrections ($\Delta\alpha_{\rm OX,corr}\approx-0.5$ to $-0.3$; e.g., 
\citealt{Gallagher2006}). The nature of the
X-ray weakness for such objects is uncertain.
It is possible that they are heavily obscured, and 
the observed \hbox{X-ray} spectra are dominated by the 
scattered/reflected component.
Alternatively,
these objects could be intrinsically \hbox{X-ray} weak compared to typical quasars,
not emitting \hbox{X-rays} at a nominal level
\citep[e.g.,][]{Gallagher2001,Leighly2007,Gibson2008,Wu2011}.
An example of a mechanism that could cause intrinsic \hbox{X-ray} weakness
is wind quenching of the accretion-disk corona, as
proposed by \citet{Proga2005}, where
the coronal X-ray emission is suppressed when the failed disk wind
falls into the corona and makes it ``too dense, too opaque, and 
consequently too cold''. 
A few BAL quasars have been suggested 
to have Compton-thick absorption
\citep[e.g.,][]{Mathur2000,Clavel2006}. These studies are
based on the weak/non detection of the source in the X-ray band, making
detailed spectral analysis infeasible. Therefore
the scenario of intrinsic \hbox{X-ray} weakness cannot be excluded.

An open question is whether even stronger \hbox{X-ray} absorption could be present
in \hbox{X-ray} weak BAL quasars,
or if instead some of these quasars 
are actually intrinsically \hbox{X-ray}
weak.
One way to address this is to observe at higher energies where
the \hbox{X-rays} are considerably more penetrating.
For example, the {\it BeppoSAX} observation of the BAL quasar
Mrk~231 revealed an absorption column density of 
$N_{\rm H}\approx2\times10^{24}$~cm$^{-2}$, with only the $\approx20$--50~keV
X-rays observed by {\it BeppoSAX} able to penetrate the 
obscuring gas \citep{Braito2004}; this discovery was later apparently
confirmed by the {\it Suzaku} data in the \hbox{15--30}~keV band 
\citep{Piconcelli2013}.\footnote{Note, 
however, that recent \nustar\ observations of Mrk~231 are not 
consistent with the {\it BeppoSAX} results, 
perhaps due to source contamination in the large beam of 
the earlier observations (S.\ Teng et al.\ 2013, in prep.).}
With the successful launch of the {\it Nuclear
Spectroscopic Telescope Array} (\nustar; \citealt{Harrison2013}) 
on 2012 Jun 13, it 
is possible to investigate this question systematically. 
\nustar\ is the first focusing telescope in orbit observing in the hard X-ray
\hbox{(3--79~keV)} band; it provides about two orders-of-magnitude
improvement in $>10$~keV sensitivity over previous hard \hbox{X-ray} missions,
as well as accurate source 
positions ($\la5\arcsec$). 
A \nustar\ survey of significantly \hbox{X-ray} 
weak BAL quasars can detect \hbox{X-rays} penetrating
the absorber at
$>10$~keV unless the absorber is very Compton-thick, 
and thereby better constrain the nature of the shielding gas and 
the disk-wind mechanism.

\subsection{The Two Targeted Broad Absorption Line Quasars}

\begin{figure*}
\centerline{
\includegraphics[scale=0.5]{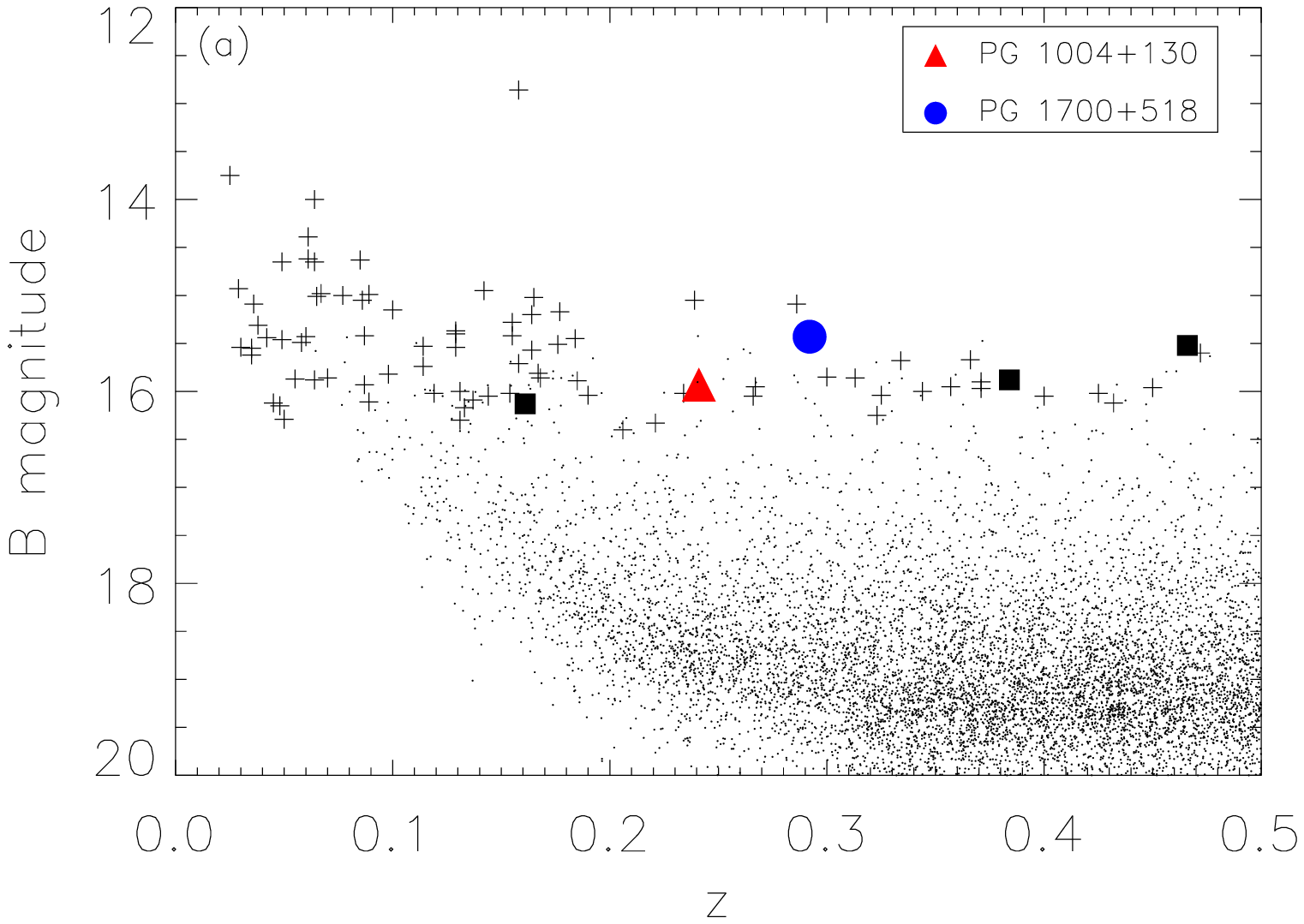}
\includegraphics[scale=0.5]{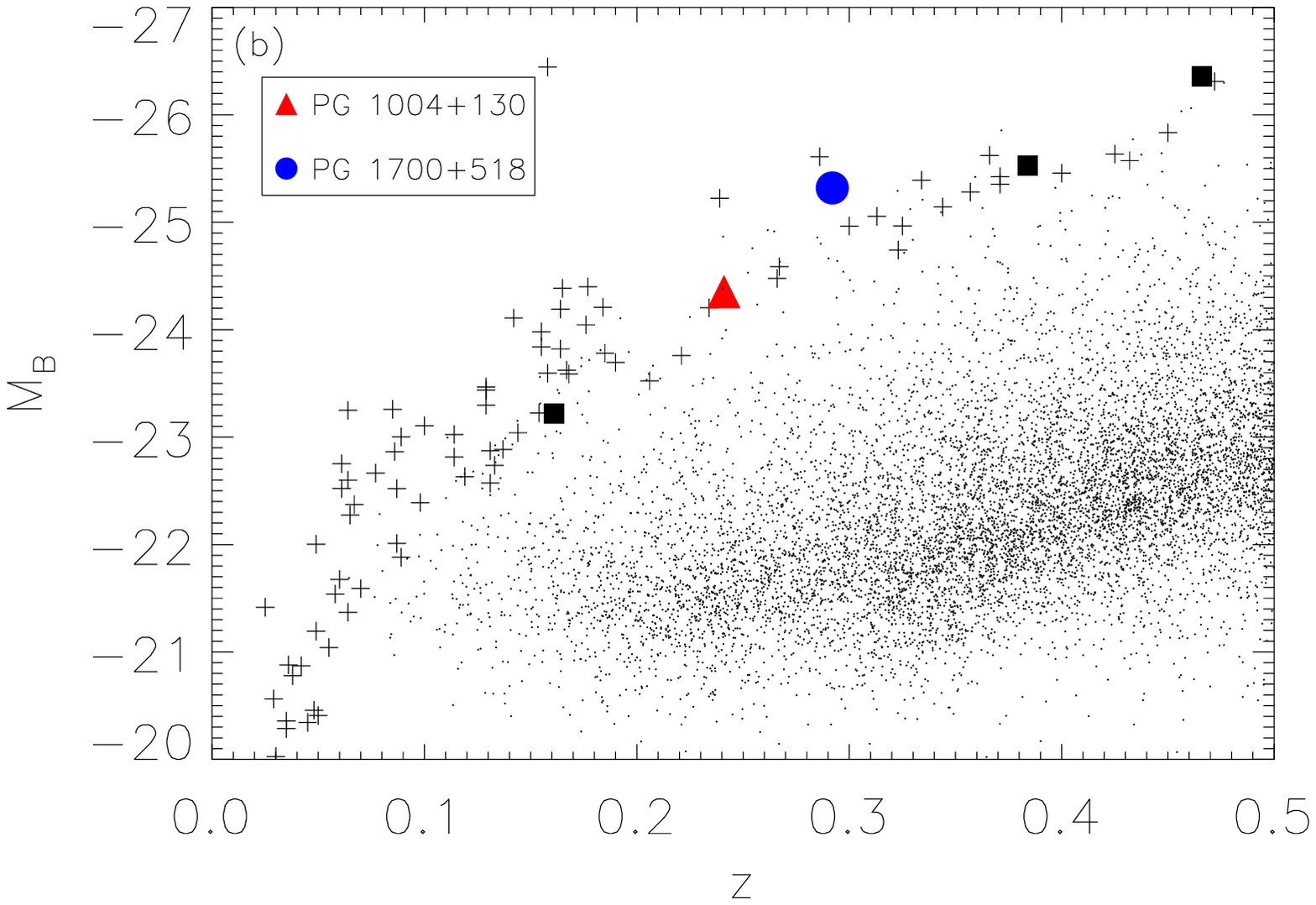}
}
\caption{
Redshift vs. (a) apparent and (b) absolute $B$-band magnitudes
for the 87 PG quasars (plus signs)
at $z<0.5$ from \citet{Schmidt1983}. 
The red triangle and blue 
filled circle represent PG~1004+130 and PG~1700+518,
respectively.
The three black squares represent 
the three additional BAL quasars in this sample.
The underlying black dots are objects from the SDSS DR7 quasar
catalog.
The $B$-band
magnitudes of the SDSS quasars were converted from the $g$-band
magnitudes, assuming an optical power-law slope of $\alpha_{\rm o}=-0.5$
($f_\nu\propto \nu^{\alpha}$; e.g., \citealt{Vandenberk2001}). The
$K$-corrections were performed assuming the same optical power-law slope.
PG~1004+130 and PG~1700+518 are among the 
optically brightest and most luminous
BAL quasars known at low redshift.
\label{fig-lz}}
\end{figure*}

As a pilot program, we selected two well studied
BAL quasars, PG~1004+130 and PG~1700+518, 
that appear \hbox{X-ray} weak and observed them with \nustar.
We selected these two targets from the 87 Palomar-Green 
(PG) quasars \citep{Schmidt1983} at $z<0.5$. These PG quasars
represent one of the best-studied samples of luminous quasars in 
the nearby universe, and the more luminous PG quasars are 
also representative local counterparts
of quasars (including BAL quasars) 
at higher redshifts (e.g., $z\approx1.5$) from
the Sloan Digital Sky Survey (SDSS; \citealt{York2000}) in terms of
luminosity.
Five BAL quasars have been classified within this PG quasar sample
\citep[see Footnote~4 of][]{Brandt2000}, two of which (PG~1001+054 and
PG~2112+059)
show soft \hbox{X-ray} weakness that was considered to be accounted for by 
moderate absorption 
\citep[e.g.,][]{Gallagher2001,Gallagher2004,Schartel2005}.\footnote{We later 
examined the {\it XMM-Newton} data for PG~1001+054, and found that 
the ionized absorption derived from spectral fitting \citep{Schartel2005}
is not sufficient to explain the X-ray weakness (still a factor $\approx7$ weaker
after absorption correction).} The other
three are also \hbox{X-ray} weak but the nature of their
X-ray weakness is not as well understood;
among these three, we chose the two lower redshift objects
as our targets here (PG~1004+130 at $z=0.241$ and PG~1700+518 at 
$z=0.292$; the other object is PG~0043+039 at $z=0.384$). 
PG~1004+130 and PG~1700+518 are among the optically brightest 
BAL quasars known; see Figure~\ref{fig-lz}.
Also shown in Figure~\ref{fig-lz} are the 87 PG quasars and
the $z<0.5$ SDSS
Data Release 7 (DR7; \citealt{Abazajian2009}) quasars from the catalog in
\citet{Schneider2010}.

PG~1004+130 and PG~1700+518 show clearly detected BALs 
\citep[e.g.,][]{Pettini1985,Wills1999,Brandt2000,Young2007}. Aside
from dust reddening in PG~1700+518,
their optical and UV spectra (continua and emission lines, excluding
the BAL regions) appear normal compared to typical quasars. 
PG~1700+518 is also sub-classified as a
low-ionization BAL (LoBAL) quasar.\footnote{
BAL quasars are broadly classified as high-ionization BAL (HiBAL)
and LoBAL quasars.
LoBAL quasars
are a subset ($\approx10\%$) of BAL quasars that have BALs 
from ions at lower ionization states such as 
\ion{Mg}{2} or \ion{Al}{3} \citep[e.g.,][]{Weymann1991,Sprayberry1992}. 
They often show signs
of dust reddening and are \hbox{X-ray} weaker than HiBAL quasars
\citep[e.g.,][]{Green2001,Gallagher2006,Gibson2009}. \label{footnote-bal}}
Both objects
are well studied and have superb multiwavelength coverage 
\citep[e.g.,][]{Ogle1999,Schmidt1999,Wills1999,Brandt2000,Miller2006,Young2007}.
PG~1004+130 is radio loud with a radio-loudness parameter
\hbox{$R\approx210$} ($R=f_{5~{\rm GHz}}/f_{\rm 4400~{\textup{\AA}}}$);
such radio-loud BAL quasars are relatively rare 
\citep[e.g.,][]{Shankar2008,Miller2009}.
PG~1700+518
is radio quiet ($R\le10$).

In the \hbox{X-ray} band,
PG~1004+130 and PG~1700+518 do not show the expected level of 
X-ray emission (determined using the 
\hbox{$\alpha_{\rm OX}$--$L_{\rm 2500~\AA}$} relation) 
for a luminous quasar at energies 
below $\approx7$--10~keV.
The \chandra\ and 
{\it XMM-Newton} spectra of PG~1004+130 show modest \hbox{X-ray} 
absorption ($N_{\rm H}\la10^{22}$~cm$^{-2}$) with a
partial-covering absorbed power-law model, and the derived photon index 
is consistent with 
the typical value for radio-loud quasars ($\Gamma\approx1.55$;
e.g., \citealt{Page2005}); however, the X-ray
continuum flux after correction for this absorption
is still $\approx11$ times lower than 
that expected from its optical/UV flux \citep{Miller2006,Miller2011}. 
PG~1700+518 has been observed by {\it XMM-Newton} with a
$\approx60$~ks exposure. The spectrum is flat ($\Gamma\approx0.2$) compared
to the typical photon index of $\Gamma\approx1.8$ for radio-quiet quasars
\citep[e.g.,][]{Reeves1997,Page2005,Just2007,Shemmer2008,Scott2011}, and
the absorption column density
was constrained to be $\approx2\times10^{23}$~cm$^{-2}$  
with an absorbed power-law model
\citep{Ballo2011}. After
correction for this strong absorption, 
PG~1700+518 is still $\approx12$ times X-ray
weaker than expected.
It was also weakly detected in a $\approx7$~ks exposure by \chandra\
($\approx14$ counts in the 0.5--8~keV band; \citealt{Saez2012}).
Note that in these $\alpha_{\rm OX}$ calculations 
the optical/UV fluxes of these two quasars have not been
corrected for any intrinsic reddening, which would render them
even X-ray weaker.

As mentioned in Section~1.1 above, absorption column densities 
constrained by \hbox{X-ray} data below $\approx7$--10~keV could be biased
for objects that are heavily obscured or even Compton-thick,
as the observed \hbox{X-ray} spectra are probably dominated by the 
scattered/reflected component.
Given the expected underlying 
X-ray continua assuming normal quasar SEDs for PG~1004+130 and PG~1700+518
(i.e., the underlying 2~keV luminosities satisfying
the \hbox{$\alpha_{\rm OX}$--$L_{\rm 2500~\AA}$} relation), 
if we were able to 
detect direct nuclear hard \hbox{X-rays} with \nustar, 
we would expect to obtain $\approx400$-600 hard X-ray ($\ga10$~keV) counts. 
Therefore, PG~1004+130 and PG~1700+518 are ideal targets
for an initial sampling of the hard \hbox{X-ray} ($>10$~keV) 
properties of \hbox{X-ray} weak BAL quasars. 

\subsection{Paper Layout and General Definitions}

This paper is organized as follows. 
In Section~2 we describe the \nustar\ observations and our 
data-analysis approach. We present photometric and spectroscopic
properties of the two targets when available.
In Section~3 we present multiwavelength properties
of the two targets, which show weak hard \hbox{X-ray} emission compared to 
typical quasars.
In Section~4 we infer the absorption
column densities from the data assuming that the weak hard 
\hbox{X-ray} emission 
is caused by obscuration by the shield gas, and
we discuss physical implications and issues related to 
the Fe~K$\alpha$ line.
We also discuss the possibility of the two targets being intrinsically
\hbox{X-ray} weak based on a
coronal-quenching model.
We summarize in Section~5.

Throughout this paper,
we use J2000.0 coordinates and a cosmology with
$H_0=70.4$~km~s$^{-1}$~Mpc$^{-1}$, $\Omega_{\rm M}=0.272$,
and $\Omega_{\Lambda}=0.728$ \citep[e.g.,][]{Komatsu2011}.
For the spectral modeling, we use the cosmic abundances of 
\citet{Anders1989} and the photoelectric absorption cross-sections
of \citet{Balucinska1992}.
We adopt the terminology that has been used in previous 
studies to describe \hbox{X-ray} weakness \citep[e.g.,][]{Laor1997,Brandt2000,Gallagher2001,Leighly2007,Gibson2008}: the term ``X-ray weak'' indicates that 
the observed \hbox{X-ray} emission is significantly weaker 
than that expected 
from the optical--UV continuum SED,
while the term 
``intrinsically \hbox{X-ray} weak'' refers to one possible cause
for the observed \hbox{X-ray} weakness where the object simply does not produce 
X-ray emission at a nominal level (one other
apparent cause would be absorption).

\section{NUSTAR OBSERVATIONS AND DATA ANALYSIS} \label{sec-data}

\subsection{\nustar\ Observations and Photometric Properties} \label{sec-pho}
\nustar\ carries two co-aligned
X-ray telescopes with a focal
length of 10.15~m focusing hard \hbox{X-ray} photons (3--79~keV) onto two
shielded focal plane modules (FPMs A and B; \citealt{Harrison2013}).
Each FPM consists of four CdZnTe pixel sensors placed in 
a two-by-two array, providing a $\approx12\arcmin\times12\arcmin$ field of view
at 10~keV.
\nustar\ has excellent angular resolution compared to previous hard
X-ray missions, with a half-power diameter (HPD) of $58\arcsec$
and a full width at half maximum (FWHM) of $18\arcsec$ independent of 
energy.

PG~1004+130 and PG~1700+518 were observed by \nustar\ with exposure times
of $32.4$~ks and $82.5$~ks, respectively. The details of the 
observations are listed in Table~\ref{tbl-obs}.
We processed the data using the \nustar\ Data Analysis Software (NuSTARDAS) 
v0.9.0 with NuSTAR CALDB 20121126. 
Cleaned calibrated event files were created using the {\sc nupipeline}
script. For each source in each of the two FPMs,
we created \hbox{X-ray} images in five bands:
\hbox{4--10~keV}, \hbox{4--20~keV}, \hbox{10--20~keV}, \hbox{20--30~keV},
and \hbox{30--79~keV}
using 
the \chandra\ Interactive Analysis
of Observations (CIAO)\footnote{See
http://cxc.harvard.edu/ciao/ for details on CIAO.}
v4.4 tool {\sc dmcopy}. The images are oversampled, and
the pixel size is $2.46\arcsec$.
We searched for sources in these images
using the CIAO tool {\sc wavdetect} \citep{Freeman2002} 
with a false-positive probability 
threshold of 10$^{-6}$ and wavelet scales of 2, 4, 8, and 16 pixels.
PG~1004+130 is relatively bright, and it is detected in multiple bands in
both FPMs. PG~1700+518 appears to be faint, and it is detected only in the 4--20~keV image of FPM~A. The background in FPM~B around the source position 
of PG~1700+518 is
$\approx20$--40\% higher than that in FPM~A at
lower energies ($\la20$~keV), rendering the source undetectable 
in this FPM.
This higher level of background in FPM~B
is caused by a larger level of stray light at
the position of the source due to unfocused
aperture leakage \citep{Harrison2013}.

\begin{figure*}
\centerline{
\includegraphics[scale=0.5]{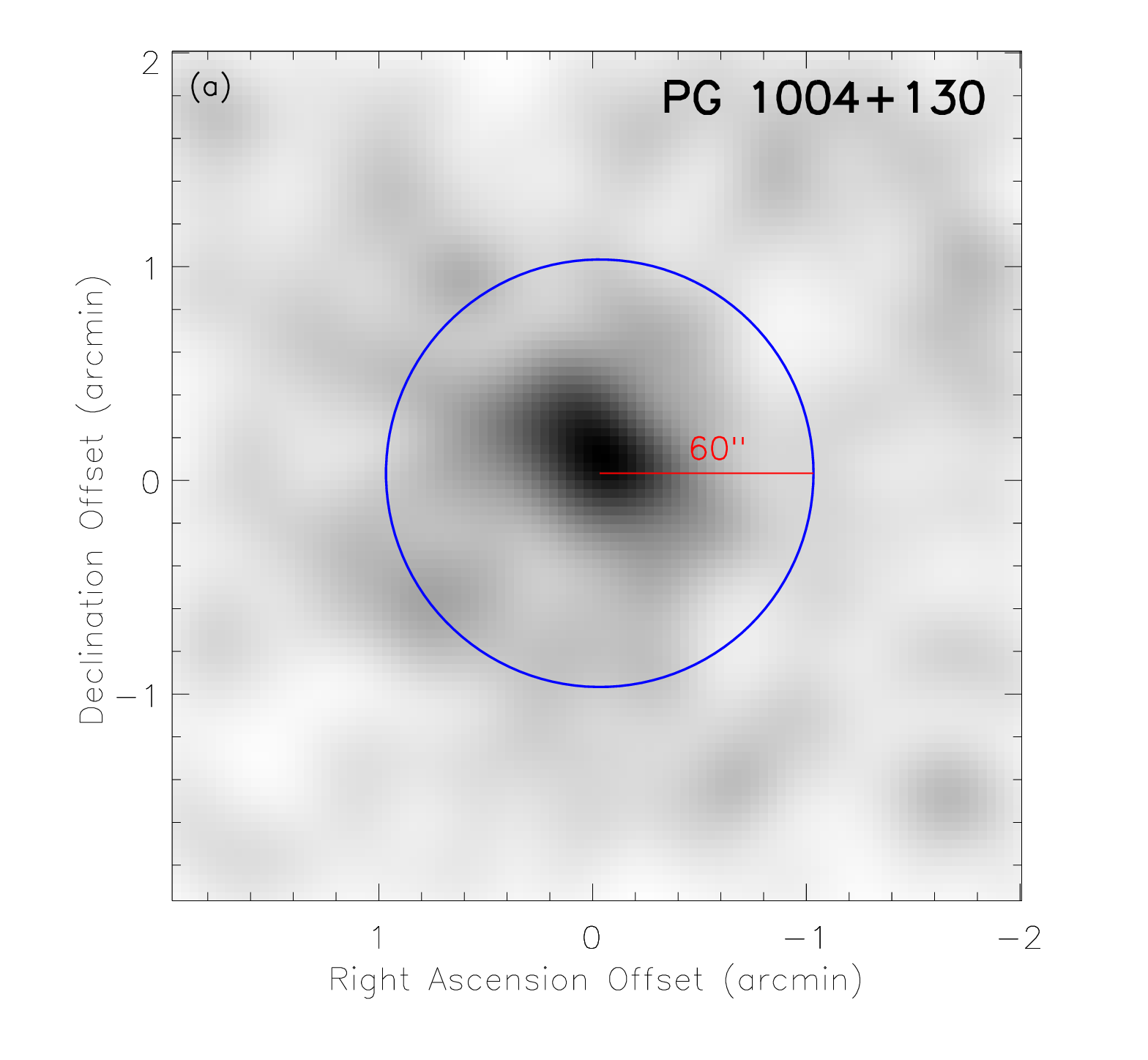}
\includegraphics[scale=0.5]{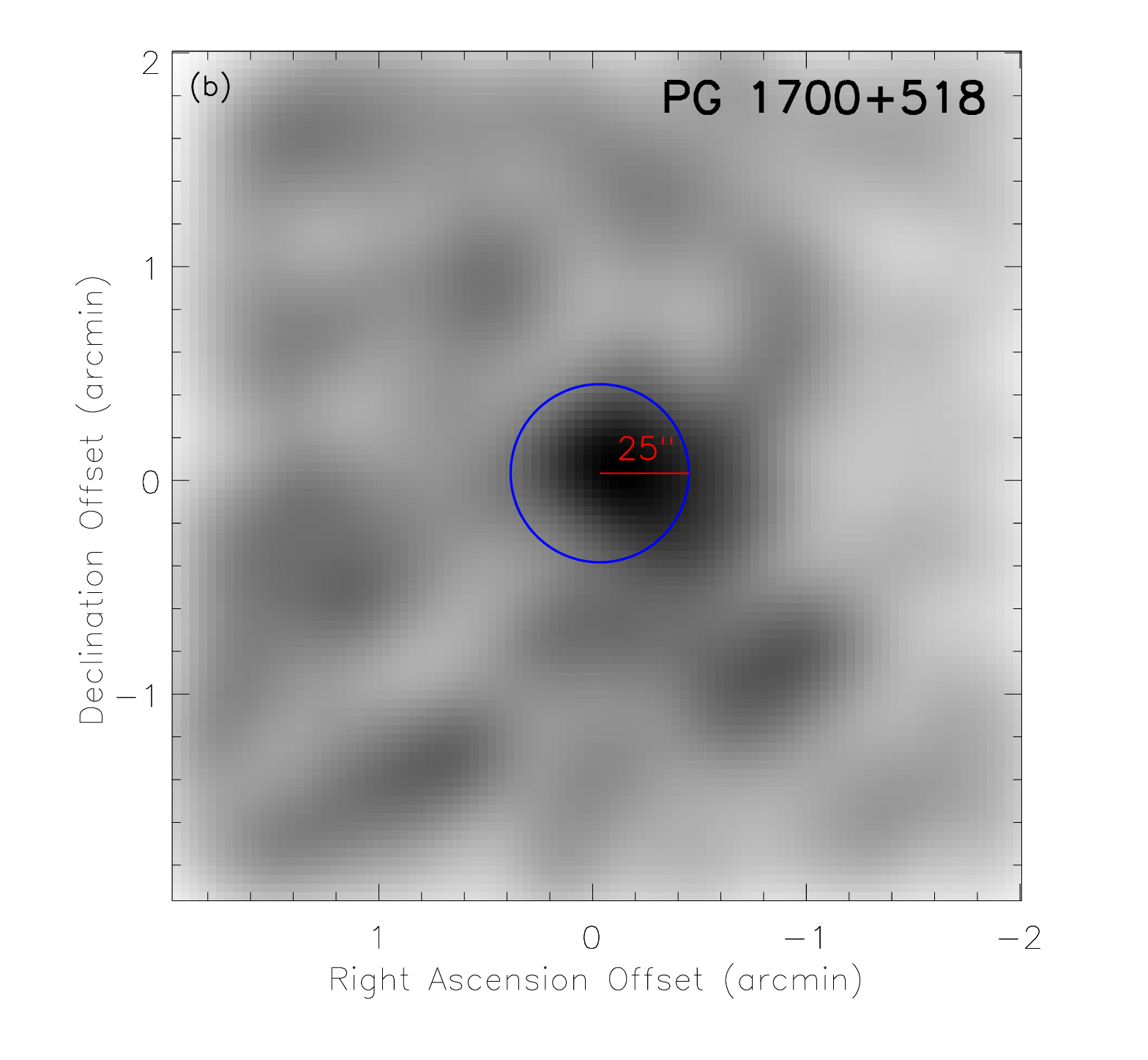}
}
\caption{
\nustar\ 4--20~keV smoothed
images of (a) PG~1004+130 and (b) PG~1700+518 in FPM~A.
Each image is $4\arcmin$ on a side, centered on the \hbox{X-ray} source
position. The images are smoothed with a Gaussian of 
width 5 pixels ($12.3\arcsec$). Image smoothing was only performed
here for display purposes, and the unsmoothed images were
used for scientific analysis.
A 60\arcsec-radius aperture was used to extract the photometric
and spectroscopic data
for PG~1004+130, and a 25\arcsec-radius aperture was used to extract the
photometric data for PG~1700+518.
\label{fig-img}}
\end{figure*}

We adopted \hbox{X-ray} positions based on the {\sc wavdetect}
detection in the 4--20~keV band, which appears to be the most
sensitive band among the five bands we studied, and the positions 
appear good upon visual inspections. 
For faint sources, \nustar\ 
provides positional accuracy to better than 5\arcsec.
The \hbox{X-ray} position
of PG~1004+130 in FPM~A is 3.1\arcsec\ away from its optical position,
and in FPM~B the offset is 0.1\arcsec.
For PG~1700+518 in FPM~A, 
the \hbox{X-ray} position is 1.5\arcsec\ away from the optical
position. 
Overall, \nustar\ provided accurate \hbox{X-ray} positions for 
these two BAL quasars, and the positional offsets are within
expectations for faint sources. This assures us that the \hbox{X-ray} 
emission detected comes from our two targets.
Neither of the two objects is detected in the \hbox{20--30~keV} or 
30--79~keV bands. More than 100 net counts would be expected for either source
in the 20--30~keV band if it had a typical 
quasar SED (i.e., a 2~keV luminosity given by
the \hbox{$\alpha_{\rm OX}$--$L_{\rm 2500~\AA}$} relation and a power-law
X-ray continuum with $\Gamma\approx1.8$). This
suggests
that their hard \hbox{X-ray} photons did not penetrate the obscuring 
material or they are intrinsically X-ray weak. No serendipitous sources were detected in the 
fields of view of
the \nustar\ observations.
The 4--20~keV images of PG~1004+130 and PG~1700+518
in FPM~A are shown in Figure~\ref{fig-img}, centered on the \hbox{X-ray} source
positions. 

We performed aperture photometry to extract source counts in the 
\nustar\ bands. Background counts were extracted
from an annular region centered on the X-ray position
with inner radius 120\arcsec\ and outer radius
180\arcsec. Different choices of the background-extraction region (e.g.,
a circular region in the source vicinity but 
outside the 120\arcsec-radius
aperture)
do not appear to affect the background estimate systematically, and 
the number of background counts generally fluctuates at the $\la10\%$ level.
For the relatively bright source, PG~1004+130, we used
a 60\arcsec-radius circular aperture to extract source counts; 
this aperture 
approximates the
$\approx88\%$ encircled-energy fraction (EEF) contour of the point
spread function (PSF).
Extended jet emission from PG~1004+130 has been detected
by \chandra\ \citep{Miller2006}, and it is included in the \nustar\ aperture 
extraction here. However, the observed X-ray flux from the extended
jet is only $\approx1\%$ of
the nuclear flux \citep{Miller2006}, and therefore it does not affect 
our analysis of the nuclear emission.\footnote{Note that the unresolved
nuclear emission
may have a more significant contribution from jet-linked X-rays created
on sub-kpc scales;
see Section~4.1.2 below.}
PG~1700+518 is only weakly detected in FPM~A, and thus
we chose a smaller source-extraction region to optimize the 
signal-to-noise ratio. We used a circular aperture 
with a radius of 25\arcsec,
corresponding to an EEF of $\approx50\%$.

For both sources in each band,
we derived their aperture-corrected net (background-subtracted)
counts. We calculated a binomial no-source probability, $P_{\rm B}$,
to assess the significance of the source
signal, defined as
\begin{equation}
P_{\rm B}(X\ge S)=\sum_{X=S}^{N}\frac{N!}{X!(N-X)!}p^X(1-p)^{N-X}~.
\end{equation}
In the above equation, $S$ is the total number of counts in the 
source-extraction region; $N=S+B_{\rm b}$, where $B_{\rm b}$
is the total number of counts in the background region; $p=1/(1+BACKSCAL)$,
where $BACKSCAL$ is the area scaling factor between the background and
source regions. $P_{\rm B}$ represents the
probability of observing the source counts by chance (due to a background 
fluctuation)
under the assumption that there is no source at the relevant location. 
It has been used to filter out low-significance sources and 
create reliable catalogs of \chandra\ sources
\citep[e.g.,][]{Broos2007,Xue2011,Luo2013}. 
For our two targets here, if the $P_{\rm B}$ value 
in a band is smaller than 0.01 ($\approx2.6\sigma$),
we considered the source to be detected and calculated
the 1$\sigma$ errors on the net counts,
which were derived from the 1$\sigma$ errors \citep{Gehrels1986}
on the extracted source and
background counts following the numerical
method in Section 1.7.3 of \citet{Lyons1991}.
If the $P_{\rm B}$ value is larger than 0.01, we considered the 
source undetected in this band and derived 
an upper limit on the source counts 
using the Bayesian approach of \citet{Kraft1991} 
for a 90\% confidence level. 
Under this criterion, PG~1004+130 and PG~1700+518 
are detected in the 
\hbox{4--10~keV}, \hbox{4--20~keV}, and \hbox{10--20~keV} bands (except 
for PG~1700+518 in FPM B).
The source counts and upper limits
of the two targets are listed in Table~2.

Using the band ratio, defined here as the
ratio between
the observed 10--20~keV and 4--10~keV counts, we derived an
effective photon index ($\Gamma_{\rm eff}$) for a power-law model with 
the Galactic absorption column density (Table~1).
We utilized the \nustar\ spectral response files (produced by the
pipeline extraction of the spectrum at the source location)
and the {\sc fakeit} command in 
XSPEC (version 12.8.0; \citealt{Arnaud1996}) to calibrate the relation
between the effective photon index and band ratio. Similarly, we calibrated a
count-rate-to-flux conversion factor that depends on the effective photon index
assuming a power-law model,
and then converted the source count rates to fluxes. 
The 1$\sigma$ error for $\Gamma_{\rm eff}$ was derived using the
errors on the counts, and flux errors were derived using the
errors on the counts and $\Gamma_{\rm eff}$.

The effective photon indices and fluxes
of the two targets are shown in Table~2. 
PG~1004+130 appears to be a fairly 
soft \hbox{X-ray} source with $\Gamma_{\rm eff}=1.7\pm0.5$,
while PG~1700+518 has a hard \hbox{X-ray} spectrum with $\Gamma_{\rm eff}=0.5\pm0.7$.
The small $\Gamma_{\rm eff}$ value (although with a large uncertainty) for
PG~1700+518
suggests that significant 
absorption ($\ga5\times10^{23}$~cm$^{-2}$) and likely also Compton 
reflection is present.
Note that a column density of $\approx2\times10^{23}$~cm$^{-2}$ 
was derived from the {\it XMM-Newton} data with an absorbed power-law model
\citep{Ballo2011}.
The \hbox{X-ray} luminosities in the 4--20~keV band (listed in Table~2) for the two 
objects are $5.3\times10^{43}$~\lum\ and $2.3\times10^{43}$~\lum, smaller
than expectations for typical quasars ($>10^{44}$~\lum).
The photometric properties
of PG~1004+130 in FPMs A and B appear consistent.
For PG~1700+518, the upper limits on the counts and fluxes
in FPM~B are consistent with those measurements in FPM~A.

\subsection{\nustar\ Spectral Analysis for PG~1004+130} \label{sec-spec}

Spectral analysis for the \nustar\ spectra of PG~1700+518
is not feasible as the extracted spectra are dominated by 
background. However, we were able to perform basic spectral analysis for 
PG~1004+130.
We extracted spectra of PG~1004+130 in FPMs A and B 
using the NuSTARDAS script {\sc nuproducts}. The same source- and 
background-extraction regions as 
used for the photometry above were adopted, and 
PSF corrections have been applied to the Auxiliary Response Files (ARFs). 
To extend the spectral coverage to lower energies ($<3$~keV),
we fit the \nustar\ data jointly with a \chandra\ spectrum. The 
41.6~ks \chandra\ observation of PG~1004+130 
with the S3 CCD of
the Advanced CCD Imaging
Spectrometer (ACIS; \citealt{Garmire2003}) was described 
in detail in \citet{Miller2006}. 
We extracted the \chandra\ spectrum using 
the CIAO tool {\sc specextract}, with a circular source 
aperture of 4\arcsec\ in
radius and a source-free
background annulus of 12\arcsec--20\arcsec\ in radius.
All the spectra were grouped with a signal-to-noise ratio of 5, and we fit the 0.3--8~keV 
\chandra\ and 3--20~keV \nustar\ spectra together with a partial-covering
absorber model as suggested by \citet{Miller2006}. 
The XSPEC model used
was {\sc zpcfabs*zpowerlw*wabs}, where {\sc zpcfabs} is a 
partial-covering absorption model, {\sc zpowerlw} is an underlying 
power-law spectrum, and {\sc wabs} is to account for 
Galactic absorption.
The spectral-shape parameters for the 
model (absorption column density $N_{\rm H}$, covering factor
$C$, and
photon index $\Gamma$) were free to vary but tied for 
the three spectra, and we let the normalization parameters
for the three spectra vary to allow for possible flux variation and 
cross-calibration uncertainties.

\begin{figure}
\centerline{
\includegraphics[scale=0.5]{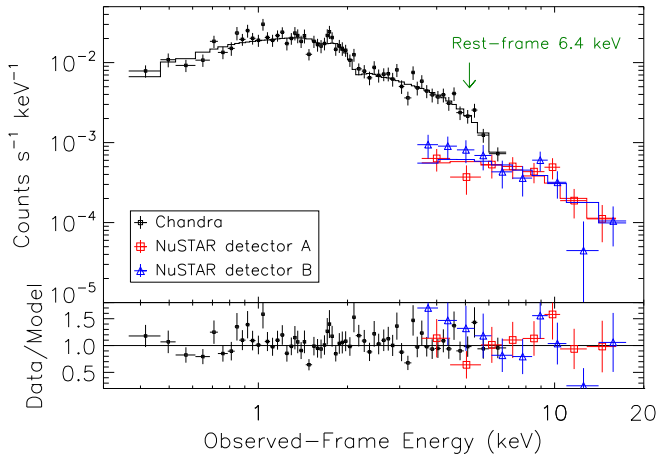}
}
\caption{
X-ray spectra of PG~1004+130 overlaid with the best-fit model.
The bottom panel shows the data-to-model ratio.
Black data points represent the \chandra\ spectrum. Red and blue 
data points represent the \nustar\ spectra in FPMs A and B,
respectively.
The spectra are fitted with a partial-covering
absorber model modified by Galactic absorption (see Section~\ref{sec-spec}
for details).
There is no apparent Fe K$\alpha$ line emission at 
rest-frame 6.4~keV (marked by the green arrow)
as is typically seen in a reflection-dominated spectrum.
\label{fig-spec}}
\end{figure}

The joint spectra overlaid with the best-fit model are displayed in 
Figure~\ref{fig-spec}. 
The best-fit model is statistically acceptable, with a null-hypothesis probability of 0.39 ($\chi^2/dof=76.9/74$). The model parameters are 
$N_{\rm H}=(1.8\pm0.6)\times10^{22}$~cm$^{-2}$, $C=0.64_{-0.15}^{+0.10}$,
and $\Gamma=1.57\pm0.19$; the quoted errors are at the 90\% confidence level
for one parameter of interest ($\Delta\chi^2=2.71$). These parameters are comparable to those
derived in \citet{Miller2006} using the \chandra\ data alone.
The intrinsic photon index, $\Gamma=1.57\pm0.19$,
is consistent with the $\Gamma\approx1.55$ typical of radio-loud quasars
\citep[e.g.,][]{Page2005}. 
We also tried to fit only the hard \hbox{X-ray} data (4--20~keV) with 
a simple power-law model modified by Galactic absorption.
The derived photon index is $\Gamma=1.7\pm0.4$ for either the 
\nustar\ data alone or the \nustar\ plus \chandra\ data set,
consistent with the $\Gamma_{\rm eff}$ value estimated from the band ratio
for PG~1004+130 (Table 2). The apparent moderate absorption and 
soft spectral shape of PG~1004+130 do not suggest Compton-thick
absorption. However, the apparent absorption is not sufficient 
to explain the \hbox{X-ray} weakness of this object, and
a Compton-thick
absorber may still be present,
if the observed spectra are dominated by a fraction 
of the jet X-ray 
emission that is not obscured by the absorber (see the further 
discussion in
Section~\ref{sec-physicalimp} below).

In the best-fit partial-covering
absorber model, the normalization parameters for 
the \nustar\ spectra in FPMs A and B are only 42\% and 46\%
of that for the \chandra\ spectrum, indicating that the 3--8~keV flux 
of PG~1004+130 observed by \nustar\ 
has dropped by a factor of $\approx2.3$ compared to the \chandra\ observation
in 2005. 
The \nustar\ fluxes derived from the
best-fit model are consistent with those derived from the
photometric approach above (Table~2) within the 1$\sigma$ errors, 
suggesting that the flux discrepancy is not likely caused by uncertainties 
introduced during the XSPEC fitting.
There was a simultaneous {\it Swift} XRT observation
of PG~1004+130 during the \nustar\ observation. However, the XRT exposure
is only 2.0~ks, and PG~1004+130 is not detected. The upper limit on the XRT
flux does not provide useful constraints on the \hbox{X-ray} variability.
PG~1004+130 is known to be variable in the \hbox{X-ray} band. The 2--8~keV \chandra\
flux is $\approx1.4$ times the {\it XMM-Newton} flux observed in
2003, and it is $\approx2.7$ times the 2--8~keV flux limit inferred from the 
1980 {\it Einstein} 0.5--4.5~keV nondetection \citep{Elvis1984,Miller2006}.
Therefore, it is likely that the \hbox{X-ray} flux of PG~1004+130
has decreased by a factor of $\approx2.3$ 
in the 2012 \nustar\ observation compared to its \chandra\ flux in 2005.
We caution that variability between 
\nustar\ and \chandra\ observations 
might also affect the best-fitting model above, since we tied the 
spectral-shape parameters in the modeling and spectral 
variability has been observed in several BAL
quasars (although it is not well constrained;
e.g., \citealt{Gallagher2004,Saez2012}).

We note that there is no apparent Fe K$\alpha$ line emission at 
rest-frame 6.4~keV (5.2~keV in the observed frame)
shown in the spectra in Figure~\ref{fig-spec}, 
as has been generally observed in \hbox{X-ray} spectra 
dominated by a reflection component \citep[e.g.,][]{Turner1997,Bassani1999,Comastri2004,LaMassa2011}.
Adding a narrow
line at 6.4~keV with a fixed width of 0.01~keV does not improve
the fit. The 90\% confidence-level upper
limit on the rest-frame Fe K$\alpha$ 
line equivalent width (EW) is $\approx178$~eV.
There is no evidence for a He-like or H-like Fe K$\alpha$
line at 6.7~keV or 7.0~keV either.

\section{SPECTRAL ENERGY DISTRIBUTIONS} \label{sec-sed}
\begin{figure*}
\centerline{
\includegraphics[scale=0.5]{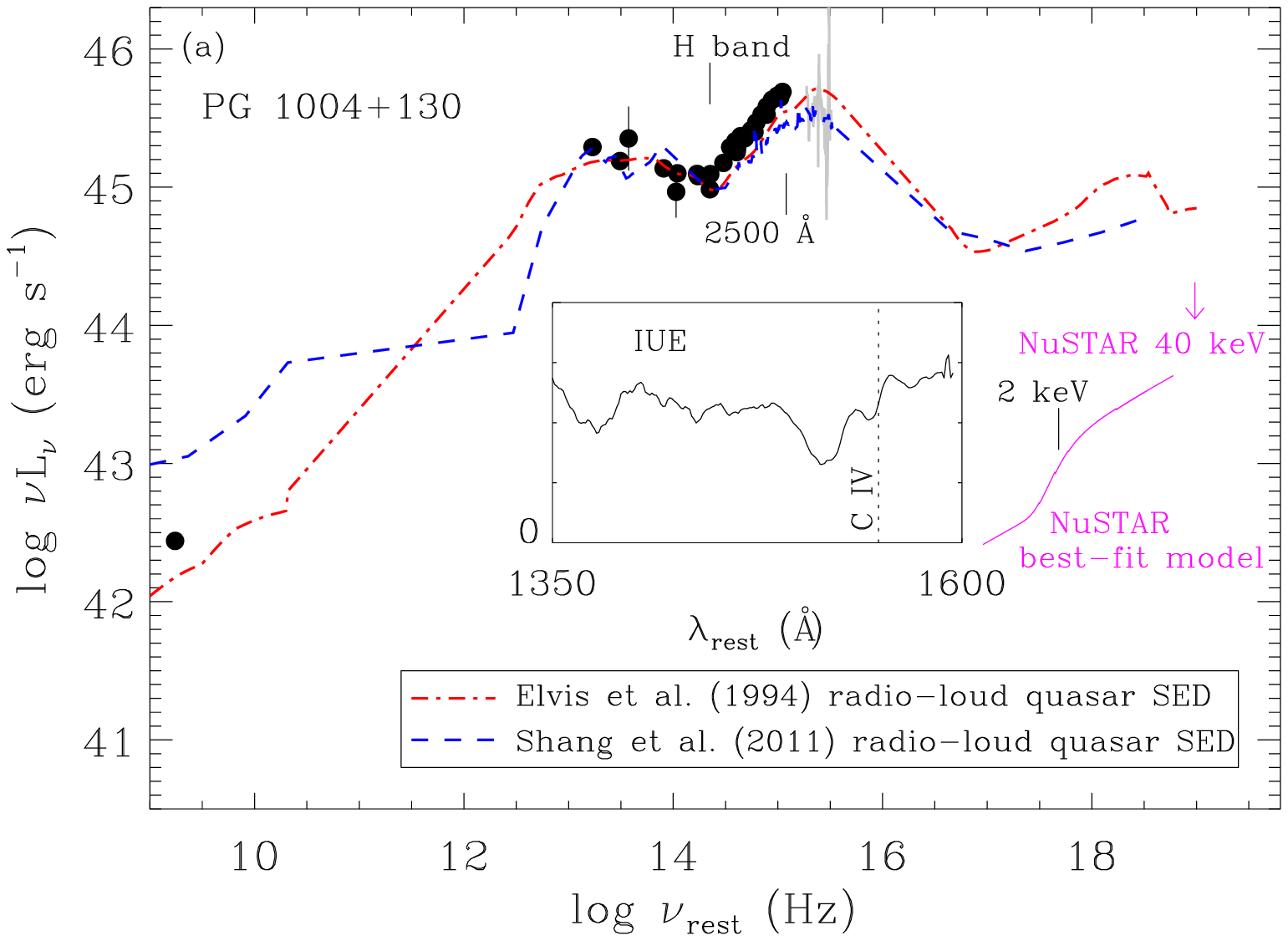}
\includegraphics[scale=0.5]{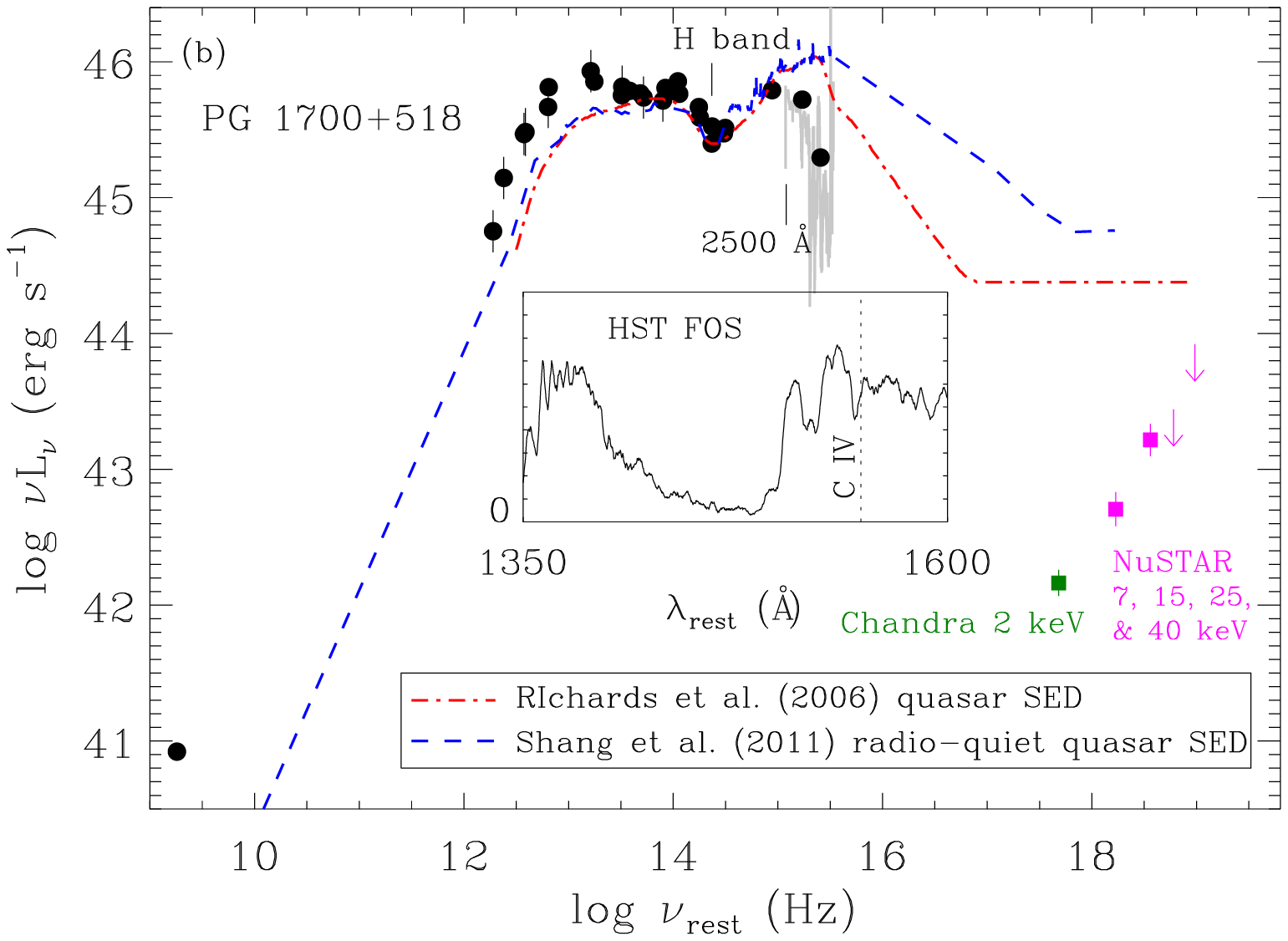}
}
\caption{
Radio through \hbox{X-ray} SEDs of (a) PG~1004+130 and (b) PG~1700+518
in the rest frame.
The black data points are from the literature and public
catalogs.
For PG~1004+130, the gray curves show the {\it HST} GHRS and {\it IUE} 
spectra;
the \hbox{best-fit} \hbox{X-ray} spectral model
(see Section~\ref{sec-spec}; 
with the \nustar\ normalization) is shown as the magenta curve, and
also shown is the upper limit on the 40~keV \nustar\ luminosity.
For PG~1700+518, the gray curve shows the {\it HST} FOS spectrum; 
the green and magenta data points/arrows are the 2~keV \chandra\ 
luminosity, 7~keV and 15~keV \nustar\ luminosities,
and 25~keV and 40~keV \nustar\ luminosity upper limits.
The optical and UV data have
been corrected for Galactic extinction.
The insets show the \ion{C}{4}~$\lambda1549$ BAL feature.
In both panels, the red dash-dotted and blue dashed
curves show the composite quasar SEDs \citep{Elvis1994,Richards2006,Shang2011}
normalized to the luminosity at the $H$ band, respectively.
The \citet{Richards2006} SED has been extrapolated to hard X-rays
assuming a $\Gamma=2$ power law.
A few strong emission lines in the \citet{Shang2011}
composite SED have been removed for display purposes.
The normalization point at 1~$\mu$m, as well as the 
2500~\AA\ and 2~keV SED points used for $\alpha_{\rm OX}$
calculations, are marked in the plots. The hard X-ray luminosities 
of PG~1004+130 and PG~1700+518 probed by \nustar\ are significantly
lower than those expected from their IR--optical SEDs.
\label{fig-sed}}
\end{figure*}

We constructed radio-to-X-ray SEDs for PG~1004+130 and PG~1700+518,
which have superb multiwavelength coverage.
We collected photometric data from the literature 
\citep{Neugebauer1979,Neugebauer1987,Schmidt1983,White1992,Haas2003,Serjeant2009},
the NRAO VLA Sky Survey (NVSS; \citealt{Condon1998}),
the {\it Wide-field Infrared Survey Explorer}
({\it WISE}; \citealt{Wright2010}), the Two Micron All Sky Survey (2MASS;
\citealt{Skrutskie2006}), the SDSS, and/or the {\it Galaxy Evolution Explorer} 
({\it GALEX}; \citealt{Martin2005}) catalogs.
We adopted the $H$ and/or $K$ band magnitudes from \citet{Guyon2006}, which
have the host-galaxy contribution removed via 
two-dimensional fitting of the images.
We also obtained {\it Hubble Space Telescope} ({\it HST}\/) GHRS and 
{\it International Ultraviolet Explorer} ({\it IUE}\/)
UV spectra for PG~1004+130 from the Mikulski Archive for Space Telescopes 
(MAST)\footnote{http://archive.stsci.edu/.}, and an {\it HST} FOS spectrum 
for PG~1700+518 from \citet{Evans2004}.
The optical and UV data have
been corrected for Galactic extinction following the dereddening 
approach presented in
\citet{Calzetti2000}.
For PG~1004+130, we adopted the \hbox{best-fit} \hbox{X-ray} spectral model
(with the \nustar\ normalization) determined in Section~\ref{sec-spec}
along with the upper limit on the 40~keV \nustar\ luminosity derived 
from the photometric information in Table~2. For PG~1700+518, we used the
2~keV \chandra\
luminosity \citep{Saez2012},\footnote{The $<2$~keV X-ray emission of 
PG~1700+518 likely has a small contribution from its star-forming activity ($<25\%$; \citealt{Ballo2011}),
which does not affect our analyses here.} 
7~keV and 15~keV \nustar\ luminosities,
and 25~keV and 40~keV \nustar\ luminosity upper limits; the \nustar\ data
are again derived from the photometric information in Table 2.
The rest-frame SEDs are shown in Figure~\ref{fig-sed}.
The \ion{C}{4}~$\lambda1549$ BAL features of these two quasars
are also shown in Figure~\ref{fig-sed} as inset panels: the broad
\ion{C}{4} troughs are clearly visible with the one in PG~1700+518 
being deeper and broader.

To compare the SEDs of these two BAL quasars to those of typical
quasars, we overlaid in Figure~\ref{fig-sed}
the composite radio-loud (for PG~1004+130) 
quasar SEDs from \citet{Elvis1994} and \citet{Shang2011}
or radio-quiet (for PG~1700+518)
quasar SEDs from \citet{Richards2006} and \citet{Shang2011}.\footnote{For 
radio-quiet quasars, the \citet{Elvis1994} sample
is biased toward X-ray bright quasars, and the \citet{Elvis2012} 
{\it XMM}-COSMOS
sample consists mainly of less-luminous AGNs, and thus these
composite SEDs were not adopted here; using these composite SEDs
would make the X-ray emission of PG~1700+518 even weaker compared
to the quasar samples. 
The \citet{Shang2011}
radio-quiet SED is also biased toward X-ray bright quasars, although not
as significantly as the one in \citet{Elvis1994}. The X-ray continuum
in the \citet{Richards2006} composite SED was derived from
the $\alpha_{\rm OX}$--$L_{\rm 2500~\AA}$ relation 
and is thus consistent with our 
interpretation of the underlying quasar X-ray spectra here.}
We extrapolated the \citet{Richards2006} 
SED to higher energies assuming a $\Gamma=2$ power law
to provide a fair comparison with the observed \nustar\ data;
studies of high-redshift quasars \citep[e.g.,][]{Page2005,Shemmer2008} indicate that 
their rest-frame $\approx20$--40 keV continua generally follow such 
a power law.
The composite
SEDs were normalized to the observed SEDs at the $H$ band,
where the SEDs are largely free of intrinsic reddening and the data have
been corrected for host-galaxy contamination \citep{Guyon2006}.
The composite quasar SEDs from different studies agree well in general
except that the \citet{Shang2011} radio-quiet SED is biased toward 
X-ray bright quasars.
From the radio to the UV,
the continuum SEDs of our two targets agree well with the 
composite SEDs. The SED of PG~1700+518 shows
intrinsic reddening in the optical and UV, which is consistent
with previous findings that BAL quasars, and especially
LoBAL quasars,
are in general redder than non-BAL quasars 
\citep[e.g.,][]{Brotherton2001,Trump2006,Gallagher2007,Gibson2009}.
By comparing the PG~1700+518 SED to the composite SEDs, we 
estimated the reddening to be $E(B-V)\approx0.14$, in agreement with 
the average value for LoBAL 
quasars \citep[e.g.,][]{Gibson2009}.

\begin{figure}
\centerline{
\includegraphics[scale=0.5]{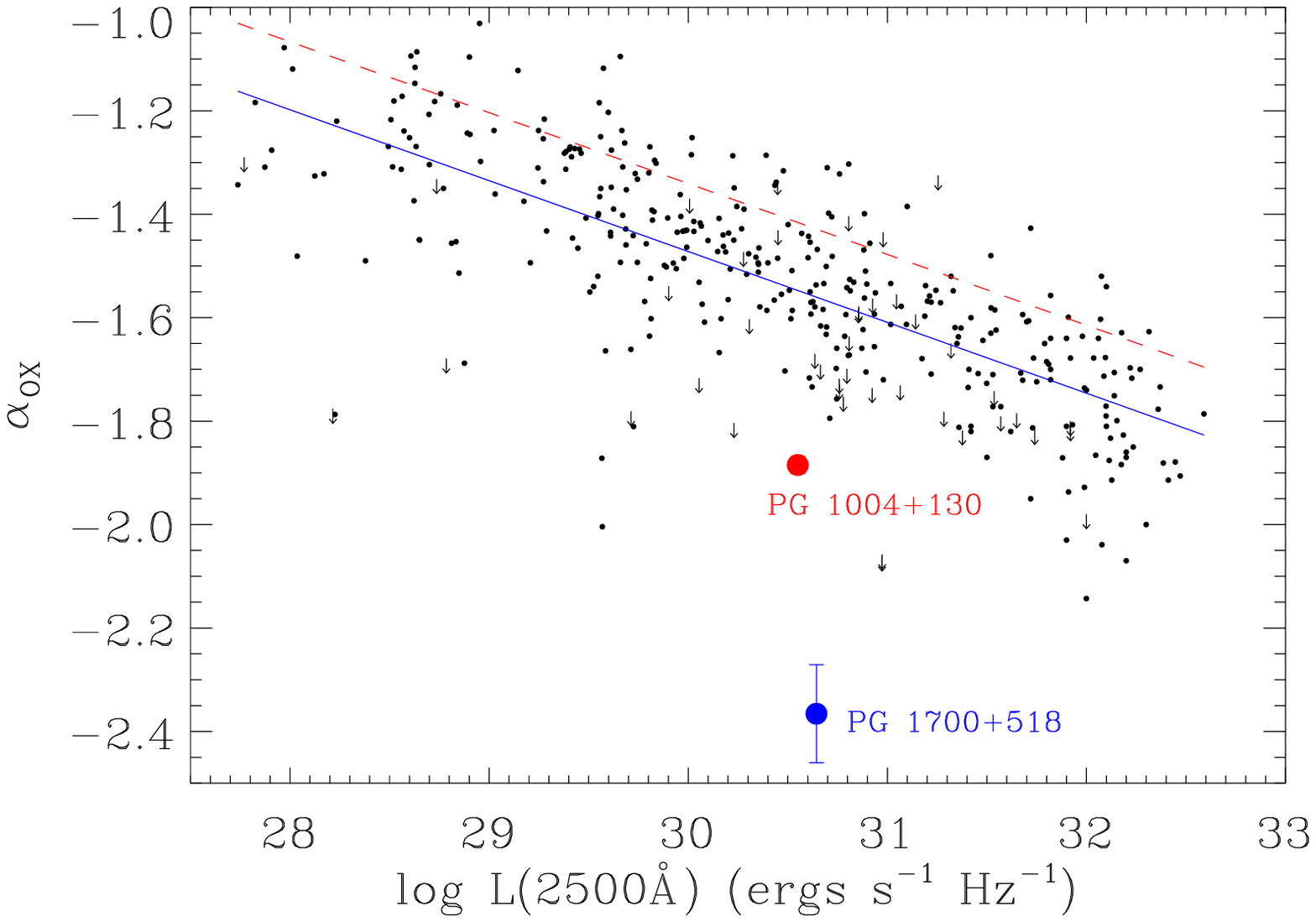}
}
\caption{X-ray-to-optical power-law slope vs. 2500 \AA\ monochromatic luminosity
for PG~1004+130 and PG~1700+518.
The small dots and downward arrows
(upper limits) are from the sample 
of \citet{Steffen2006} with the solid blue line showing the
$\alpha_{\rm OX}$--$L_{\rm 2500~\AA}$ relation.
The dashed red line shows the \citet{Steffen2006} relation modified with
the excess \hbox{X-ray} luminosity expected for the radio loudnesss of PG~1004+130 
(derived based on Section~4 of \citealt{Miller2011}). 
The errors on $\alpha_{\rm OX}$ for the two BAL quasars 
were propagated from the errors on the \hbox{X-ray} fluxes; 
it is smaller than the symbol size and is thus not visible for PG~1004+130.
PG~1004+130 and PG~1700+518 are $\approx16$ and $\approx120$
times weaker than expected in the soft X-rays ($\approx2$~keV), respectively.
\label{fig-aox}}
\end{figure}

Both BAL quasars are significantly \hbox{X-ray} weak compared to
typical quasars. 
We computed the $\alpha_{\rm OX}$ parameter, defined as
$\alpha_{\rm OX}=-0.3838\log(f_{2500~{\rm \AA}}/f_{2~{\rm~keV}})$,
for quantitative comparison.
The rest-frame 2500~\AA\ flux density ($f_{2500~{\rm \AA}}$) 
was determined by 
interpolating/extrapolating the \hbox{optical--UV} photometric data points.
For PG~1700+518, the dust reddening was not corrected; reddening correction 
would 
increase the 2500~\AA\ flux density by $\approx50\%$ (i.e.,
more negative $\alpha_{\rm OX}$).
The rest-frame 2~keV flux density was derived from the best-fit 
spectral model (see Section~\ref{sec-spec} above) for PG~1004+130,
and it was adopted from the weighted-average value 
of \chandra\ and {\it XMM-Newton} observations in
\citet{Ballo2011} and \citet{Saez2012} for PG~1700+518. 
The $\alpha_{\rm OX}$ values
are $-1.88\pm0.02$ and $-2.36\pm0.09$ for PG~1004+130 and~PG~1700+518, respectively. The $\alpha_{\rm OX}$--$L_{\rm 2500~\AA}$ plot is shown in Figure~\ref{fig-aox}.

For comparison, the \citet{Steffen2006} 
\hbox{$\alpha_{\rm OX}$--$L_{\rm 2500~\AA}$} relation
for radio-quiet quasars predicts $\alpha_{\rm OX}=-1.56\pm0.20$
for PG~1700+518. PG~1004+130 is radio loud, and the \citet{Steffen2006}
relation is not applicable. We thus adjusted the expected $\alpha_{\rm OX}$ value
by accounting for the excess \hbox{X-ray} luminosity
expected for a radio-loud quasar with the radio loudness of PG~1004+130 
(see the relation in Section 4 of \citealt{Miller2011}).
The resulting expected $\alpha_{\rm OX}$ value for 
PG~1004+130 is $-1.42\pm0.26$.
Therefore, although the continuum radio-to-UV SEDs of these two BAL quasars
resemble those of typical quasars, their soft \hbox{X-ray} ($2$~keV) luminosities
are $\approx16$ (for PG~1004+130, with a 1$\sigma$ range of 3--76) 
and $\approx120$ (for PG~1700+518, with a 1$\sigma$ range of 36--400) 
times lower 
than the typical values.
After
corrections for apparent \hbox{X-ray} absorption determined from $<10$~keV 
data, 
their soft \hbox{X-ray} luminosities are still 11 and 12 
times lower than expected (see Section~1.2).

It has been suggested that there is an additional correlation between 
$\alpha_{\rm OX}$ and the Eddington ratio \citep[e.g.,][]{Lusso2010},
where $\alpha_{\rm OX}$ is lower (i.e., more X-ray weak) when the
Eddington ratio is higher. However, the correlation suggests 
super-Eddington accretion for $\alpha_{\rm OX}<-1.8$ 
(also lacking sampling in this regime),
which does not appear to be 
the case for our two targets (see Section~4.1.2
for their Eddington ratios).

In the hard \hbox{X-ray} bands probed by \nustar, 
the luminosities of PG~1004+130 and PG~1700+518 are
also below expectations. At rest-frame 20~keV, the \nustar\ luminosities
are more than an order of magnitude 
lower than those of typical quasars; even at 
rest-frame 40~keV, where the \nustar\ sensitivity to faint sources is lower,
the 90\% confidence-level upper limits on the luminosities
are still a few times smaller than expectations.
If these two BAL quasars are able to produce \hbox{X-ray} emission
as typical quasars do, the \nustar\ data indicate that 
Compton-thick obscuration is present that blocks not only the soft X-rays
but also the hard X-rays.

\section{DISCUSSION}

Given the \nustar\ and multiwavelength data presented above
for the two BAL quasars targeted, it appears
that either (1) there is highly Compton-thick obscuration and even the hard 
X-rays probed by \nustar\ cannot penetrate the obscuring material, or 
(2) these
objects are intrinsically \hbox{X-ray} weak.
In the following, we discuss these two scenarios in more detail.

\subsection{Compton-thick Absorption?} \label{sec-ct}

\subsubsection{Absorption Column-Density Constraints} \label{sec-columnd}

For many BAL quasars, the observed X-ray weakness can be fully accounted
for by absorption (see Section~1.1).
The estimated absorption column densities for these objects have a continuous
distribution between $\approx10^{21}$~cm$^{-2}$ and
$\approx5\times10^{23}$~cm$^{-2}$, and thus we expect that
there should probably be objects with higher column densities (e.g., 
$10^{24}$--$10^{25}$~cm$^{-2}$) as well.
Furthermore, one of our targets, PG~1700+518, 
shows an effective
photon index of $\Gamma\approx0.5$, suggesting significant absorption.
Therefore, it is natural to consider the possibility of 
Compton-thick absorption for objects that 
show significant hard X-ray weakness.
The underlying \hbox{X-ray} continua of these objects 
would then be comparable to those of typical quasars, but
their observed X-ray spectra would probably be dominated by 
the scattered/reflected component.

In this scenario, we can estimate the expected absorption column density
for PG~1004+130 and PG~1700+518 based on
the ratio of the observed
broad-band \hbox{X-ray} flux
and the expected intrinsic 20~keV flux 
density for
a given value of $\alpha_{\rm OX}$ (cf. \citealt{Gallagher1999}); to obtain this intrinsic 20~keV flux 
density, we determined the intrinsic 2~keV flux density using $\alpha_{\rm OX}$
and extrapolated to 20~keV assuming
a power-law spectrum with $\Gamma=1.8$.\footnote{The $\alpha_{\rm OX}$ 
parameter has also been defined at energies higher than 2~keV 
(e.g., 10~keV; \citealt{Young2010}).
The $\alpha_{\rm OX}$--$L_{\rm 2500~\AA}$ relation with $\alpha_{\rm OX}$ 
defined at 2~keV has a relatively small dispersion due to the small
errors in the 2~keV flux measurements, and it is therefore adopted here. 
The intrinsic 20~keV flux density predicted with $\alpha_{\rm OX}$ defined
at 10~keV would lead to consistent results within the errors.}
The relation between this observed ratio and the column density was calibrated
using the {\sc MYTorus} model 
\citep{Murphy2009}\footnote{See \url{http://www.mytorus.com/} for details.}
implemented in XSPEC. The {\sc MYTorus} model
computes the transmitted and scattered \hbox{X-ray} spectra
from a toroidal-shaped absorber/reprocessor in a physical 
and self-consistent way,
and it was designed to model \hbox{X-ray} spectra in the Compton-thick regime.
The model was calculated for neutral material. However, for our purpose
of constraining the basic absorption column density in these BAL 
quasars, the model
is likely also applicable to ionized material, as high-energy \hbox{X-ray} attenuation
is dominated by Compton scattering (not photoelectric absorption) 
in the Compton-thick regime.
Two important geometrical parameters of the {\sc MYTorus} model are 
the \hbox{half-opening} angle of the obscuring medium and the inclination 
angle (0\degr\ corresponds to a face-on viewing angle). 
The default \hbox{half-opening} angle was set to 60\degr\ (corresponding to a
covering factor of 0.5), and we assumed an inclination
angle of 80\degr\ (large inclination angles are generally expected for BAL
quasars; see Section 1.1). We also explored the effects of different inclination
angles and a different geometry
with a \hbox{half-opening} angle of 37\degr\
(corresponding to a covering factor of 0.8). 
The other parameters of the model, 
such as the relative cross-normalization factors of
different components, were set as the default values (see Section 8.2 of the {\sc MYTorus} manual).

\begin{figure*}
\centerline{
\includegraphics[scale=0.5]{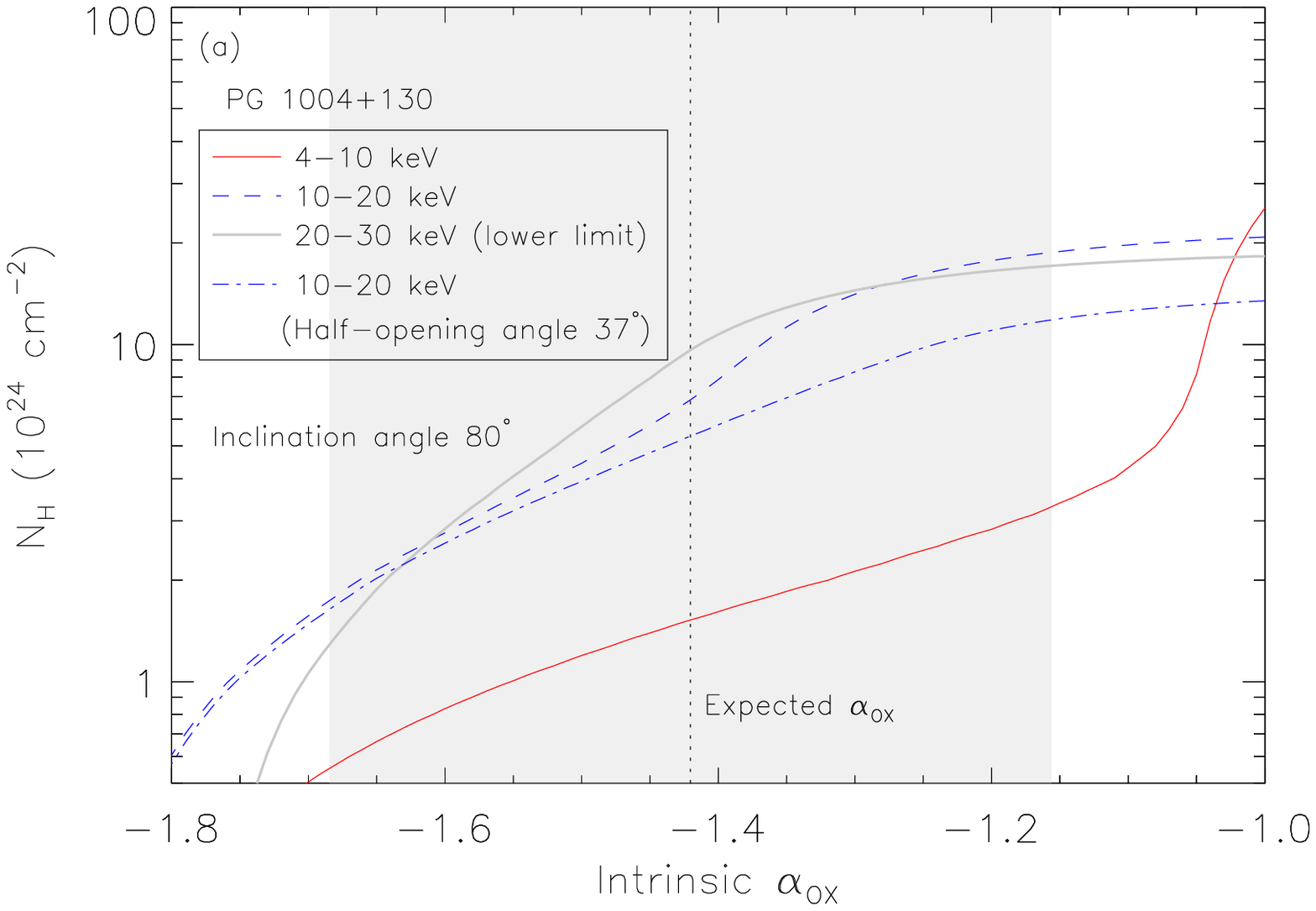}
\includegraphics[scale=0.5]{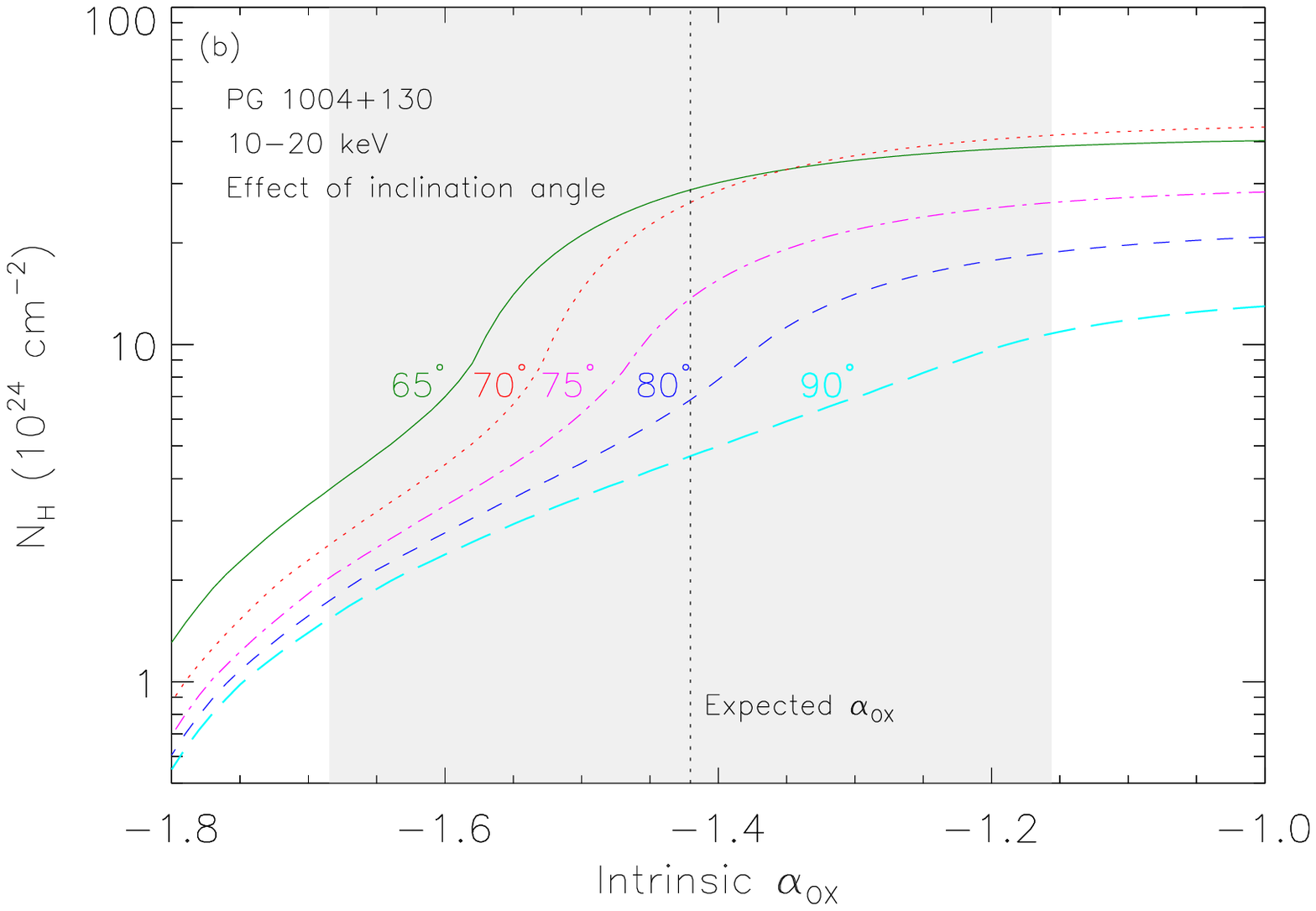}
}
\caption{
Expected column-density value as a function of the assumed intrinsic
$\alpha_{\rm OX}$ value for PG~1004+130.
The {\sc MYTorus} model was used to
determine the expected $N_{\rm H}$ value given the ratio of the observed
broad-band flux (or the flux upper limit)
and the expected intrinsic 20~keV flux
density for
a given value of $\alpha_{\rm OX}$; to obtain this intrinsic 20~keV flux
density, we determined the intrinsic 2~keV flux density using $\alpha_{\rm OX}$
and extrapolated to 20~keV assuming
a power-law spectrum with $\Gamma=1.8$.
The default {\sc MYTorus} parameters were adopted, with a 60\degr\
half-opening angle of the obscuring medium.
An inclination angle of 80\degr\ is assumed in panel (a), and
the effect of different inclination angles is explored in panel
(b).
The highest column density available in the {\sc MYTorus} model is
$10^{25}$~cm$^{-2}$, and thus any constraint above this value was
derived from extrapolation and may have a large uncertainty.
In panel (a), the red solid, blue dashed, and gray curves
indicate constraints obtained from
the 4--10~keV, 10--20~keV, and 20--30~keV photometric data; the 20--30~keV
data provide 90\% confidence-level lower limits
on the column density due to non-detection of the source.
The 10--20~keV constraints for a different geometry of
the obscuring medium (a half-opening angle of 37\degr)
are shown as the blue
dash-dotted curve.
The vertical dotted line and the shaded region represent
the expected $\alpha_{\rm OX}$ value
and its 1$\sigma$ uncertainty, which have been adjusted
based on the radio loudness of PG~1004+130
(see Section~\ref{sec-sed}).
\label{fig-1004nh}}
\end{figure*}

\begin{figure*}
\centerline{
\includegraphics[scale=0.5]{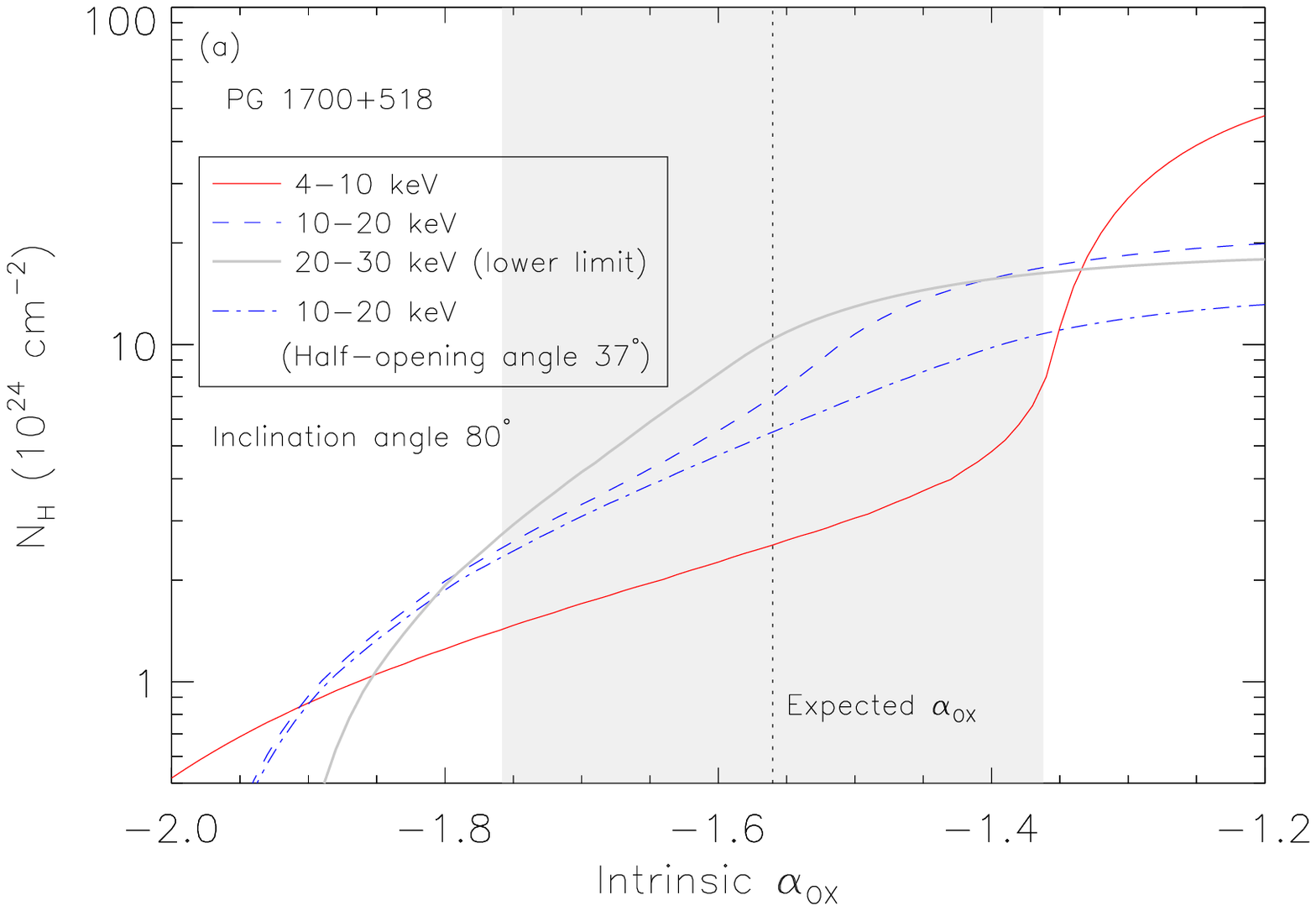}
\includegraphics[scale=0.5]{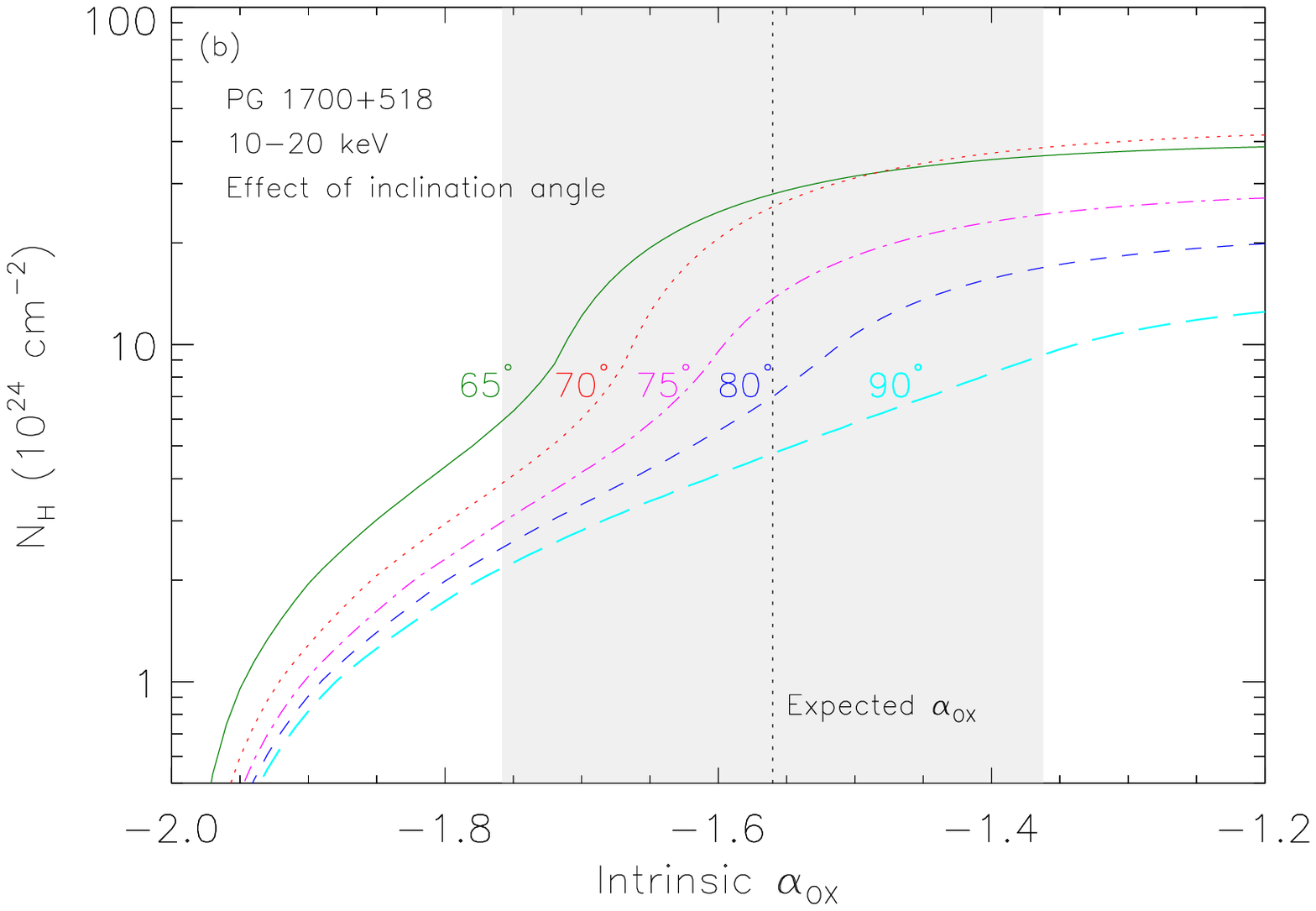}
}
\caption{
Same as Figure~\ref{fig-1004nh} but for PG~1700+518.
\label{fig-1700nh}}
\end{figure*}

We derived column-density constraints using the \nustar\ fluxes
in three bands, \hbox{4--10~keV}, 10--20~keV, and 20--30~keV (see Table 2), 
under a range
of assumed $\alpha_{\rm OX}$ values. Since neither object is
detected in the 20--30~keV band, the column densities constrained in this 
band are 90\% confidence-level lower limits. The results are displayed
in Figures~\ref{fig-1004nh} and \ref{fig-1700nh}.
The harder bands (e.g., 10--20~keV) probed by \nustar\
generally provide much tighter constraints than the softer bands (e.g., \hbox{4--10~keV}).\footnote{The column density inferred from the \hbox{4--10~keV} data
is smaller than the \hbox{10--20~keV} one, indicating that 
the X-ray weakness
is less prominent in the 4--10~keV band and the observed spectral
shape differs from the one predicted by the {\sc MYTorus} model. 
This is probably due to 
additional 4--10 keV continuum emission from a jet (for radio-loud objects)
or a scattering medium.}
The different geometric assumptions (\hbox{half-opening} angle and
inclination angle) also affect the results somewhat.
Note that the highest column density available in the {\sc MYTorus} model is 
$10^{25}$~cm$^{-2}$, and thus any constraint above this value was
derived from extrapolation and may have a large uncertainty.
Nevertheless, Compton-thick absorption appears required for any typical
assumption about the intrinsic $\alpha_{\rm OX}$ value.
For the $\alpha_{\rm OX}$ values expected from the 
$\alpha_{\rm OX}$--$L_{\rm 2500~\AA}$ relation (see Section~\ref{sec-sed}),
shown as the vertical dotted lines in Figures~\ref{fig-1004nh} and \ref{fig-1700nh},
the 10--20~keV \nustar\ data indicate $N_{\rm H}=(6.9_{-5.1}^{+11.9})\times10^{24}$~cm$^{-2}$ 
for PG~1004+130 and $N_{\rm H}=(7.0_{-4.5}^{+9.9})\times10^{24}$~cm$^{-2}$ for PG~1700+518,
for a \hbox{half-opening} angle of 60\degr\ and an inclination
angle of 80\degr. These column densities correspond 
to Thomson optical depths of $\tau_{\rm T}\approx5$. The uncertainty
of the estimated $N_{\rm H}$ was determined from the scatter in the expected
$\alpha_{\rm OX}$ value which is much more significant than 
the uncertainty of the observed 
flux. At $N_{\rm H}\approx7.0\times10^{24}$~cm$^{-2}$, the observed 
spectrum computed from the {\sc MYTorus} model is completely 
dominated by the scattered component; it appears flat with a high-energy 
hump peaking
around observed-frame 20~keV (e.g., see Figure~6.1 of the {\sc MYTorus}
manual), and the observed 10--20~keV flux is absorbed by a factor 
of $\ga10$.

The parameterization of
the {\sc MYTorus} model cannot, of course, fully reproduce the
complex absorption environments of these BAL quasars, but we consider
it to be the best available approximation for the purpose of deriving 
basic column-density constraints.
If the observed \hbox{X-ray} emission has contributions from additional 
continuum components that were not considered in the {\sc MYTorus}
modeling, our estimation of the
column density above would likely be an underestimate as the 
scattered/reflected component associated with the absorber would be weaker
than the observed emission. Examples of possible additional X-ray
components include jet emission and 
the nuclear continuum scattered by a large-scale medium; 
see Section~\ref{sec-feline} below for details.

\subsubsection{Physical Implications of Compton-thick Absorption} \label{sec-physicalimp}

In the disk-wind scenario for BAL quasars, the outflowing wind is mainly 
radiatively driven
by UV line pressure. In some models \citep[e.g.,][]{Murray1995,Proga2000},
the UV radiation originates from the center of the AGN
and is filtered by the \hbox{X-ray} shielding gas. This assumption requires
the shielding gas not be Compton thick; otherwise, the UV radiation
would also be blocked and the wind would lose its driving force. In other
models \citep[e.g.,][]{Proga2004}, this assumption
is relaxed, and the UV radiation is allowed to be produced exterior to the 
shielding gas (e.g., see Figure~\ref{fig-art}); 
our finding of potential Compton-thick absorption in these two
BAL quasars supports this latter geometry.
Given the surface-temperature distribution of a standard
accretion disk \citep{Shakura1973}, the radius of the disk region that emits
most strongly in the UV can be estimated as
\begin{equation}
R_{\rm UV}=3.2\times10^{15}\left(\frac{\eta}{0.1}\right)^{-1/3}\dot{m}_{\rm Edd}^{1/3}
\left(\frac{M_{\rm BH}}{10^8M_\sun}\right)^{2/3}\left(\frac{\lambda}{1550~{\textrm \AA}}\right)^{4/3}\ \textrm{cm},
\end{equation}
where $\eta\approx0.1$ is the accretion efficiency and $\dot{m}_{\rm Edd}$ is the
accretion rate in units of the Eddington accretion rate (Eddington ratio).
The black-hole masses ($M_{\rm BH}$) for PG~1004+130 and PG~1700+518 
are \hbox{$(1.9\pm0.4)\times10^{9}~M_\sun$} \citep{Vestergaard2006}
and \hbox{$(7.8^{+1.8}_{-1.6})\times10^{8}~M_\sun$} \citep{Peterson2004}, respectively
(the radius does not depend strongly on the black-hole mass, and
thus the uncertainty on the mass estimate does not affect
the derived compactness of the absorber significantly).
Their bolometric luminosities can be calculated using the 
normalized \citet{Elvis1994} or \citet{Richards2006}
composite SEDs in Figure~\ref{fig-sed},
and they are $2.3\times10^{46}$~\lum\ for PG~1004+130 and 
$4.2\times10^{46}$~\lum\ for PG~1700+518. 
The Eddington ratios are then $\dot{m}_{\rm Edd}=0.09$ for PG~1004+130
and $\dot{m}_{\rm Edd}=0.41$ for PG~1700+518.
The derived UV-emitting radii are $\approx20R_{\rm s}$ (\hbox{$\approx10^{16}$~cm}) 
for PG~1004+130
and $\approx40R_{\rm s}$ ($\approx10^{16}$~cm) for PG~1700+518,
where $R_{\rm s}=2GM_{\rm BH}/c^2$ is the Schwarzschild radius.
Therefore, Compton-thick 
absorption
constrains the absorbing medium (i.e., shielding gas) to be compact, 
located within $\approx10^{16}$~cm of the SMBH. 
Recent microlensing studies \citep[e.g.,][]{Pooley2007,Blackburne2011,Jimenez2012} 
suggest that the optical/UV emitting regions
of quasar accretion disks are $\approx3$--30 times larger than
those predicted from the standard \hbox{accretion-disk} model, and thus
the above constraint on the compactness of the absorbing medium
could be relaxed.

PG~1004+130 is a radio-loud quasar. Under the assumption that
its intrinsic \hbox{X-ray} emission is normal, most (\hbox{$\approx70\%$} based 
on its radio loudness) 
of its \hbox{X-ray} emission
should come from the radio jets \citep[see Section 4 of][]{Miller2011}.
The observed \hbox{X-ray} weakness thus implies that
the jet emission is also heavily absorbed. 
Based on an \hbox{X-ray} survey of 21 radio-loud BAL quasars, 
\citet{Miller2009} concluded that jet emission 
is likely partially absorbed in 
these objects. Therefore, it is possible that the underlying jet 
emission of PG~1004+130 is mostly absorbed by the same Compton-thick
material (i.e., the shielding gas) that blocks the other 
nuclear \hbox{X-ray} emission. The observed \hbox{X-ray} spectra
may even be dominated by the unobscured portion of the jet emission
if the nuclear emission is strongly absorbed. 
Such a jet-dominated \hbox{X-ray} spectrum 
can explain the 
soft spectral shape ($\Gamma=1.57$; see Section~\ref{sec-spec})
observed that is consistent with those for
radio-loud quasars, and it may also be responsible for 
the nondetection of the Fe K$\alpha$ line
(see details below).

\subsubsection{Dilution of Fe~K$\alpha$ Line Emission in PG~1004+130 
by Jet-Linked X-rays?} \label{sec-feline}
For a Compton-thick AGN with obscuration by neutral matter,
a strong narrow 
Fe~K$\alpha$ emission line at 6.4~keV with an EW of order 1--2~keV 
is expected if the continuum is reflection dominated
\citep[e.g.,][]{Ghisellini1994,Matt1996}.
A strong Fe K$\alpha$ line is observed in the majority of
the known Compton-thick AGN population 
\citep[e.g.,][]{Turner1997,Bassani1999,Comastri2004,LaMassa2011}.
There is no Fe K$\alpha$ line detected in the spectra of PG~1004+130;
the upper limit on the EW is $\approx180$~eV (see Section~\ref{sec-spec}).
For PG~1700+518, no Fe K$\alpha$ line is detected either, though
the {\it XMM-Newton} spectrum cannot constrain an upper limit
due to the dominating high background at high energies, which prevents
even a detection of the continuum in the Fe K$\alpha$ band \citep{Ballo2011}.
Therefore, PG~1004+130 appears to be an unusual Compton-thick AGN
without a strong Fe K$\alpha$ line. 

However, PG~1004+130 is a luminous radio-loud type 1 BAL quasar, and
the physical nature (e.g., location, geometry, and ionization state) 
of Compton-thick absorption in 
such objects might differ from those in local 
type 2 Seyfert galaxies. One plausible cause for 
PG~1004+130 lacking a strong Fe K$\alpha$ line is that the 
line is diluted by a jet-linked \hbox{X-ray} continuum which
could dominate over the scattered/reflected nuclear continuum (cf. \citealt{Miller2006}).
Jet dilution of the Fe K$\alpha$ line has been observed in the general population of
radio-loud AGNs \citep[e.g.,][]{Eracleous2000,Grandi2006}.
In the Compton-thick regime, 
the Fe~K$\alpha$ line flux drops rapidly (although the line EW increases)
when $N_{\rm H}$ increases,
especially at large inclination angles \citep[e.g.,][]{Yaqoob2010}. Therefore,
the EW of the Fe K$\alpha$ line could be reduced substantially by any
increase of the continuum level.
Utilizing the {\sc MYTorus} model above including the default
Fe K$\alpha$ emission-line component
and assuming $N_{\rm H}=7\times10^{24}$~cm$^{-2}$,
we estimated that $6\%$ (the fraction that would make 
PG~1004+130 $\approx16$ times 
\hbox{X-ray} weaker than expected; see Section~\ref{sec-sed}) 
of the intrinsic jet continuum
that is not absorbed could dilute the EW of
the Fe K$\alpha$ line from $\approx1.5$~keV to $\approx130$~eV.
If the covering factor of the shielding gas is smaller than the 
{\sc MYTorus} default value (0.5), which is likely the case
given the small covering factor of the disk wind ($\approx0.2$; see Section~1.1),
the EW of the Fe K$\alpha$ line could be even smaller \citep[e.g.,][]{Ikeda2009}.
We caution that with the dilution from the jet emission, 
the column density estimated in Section~\ref{sec-columnd}
is likely a lower limit.

For completeness, we mention below
a few additional possible explanations
for the lack of a strong Fe K$\alpha$ line from PG~1004+130, which
might also be applicable for other BAL quasars.

\begin{enumerate}

\item
The strong Fe K$\alpha$ line could be diluted by
the scattered continuum from a highly ionized Compton-thin medium
that surrounds the SMBH on a larger scale than the Compton-thick 
material \citep[e.g.,][]{Murphy2009b,Yaqoob2009}.\footnote{Also see the presentation at
http://cxc.harvard.edu/ChandraDecade/pro-ceedings/session\_13.html\#talk57.}
We note that the scattering geometry in the \hbox{X-ray} may
differ significantly from that studied spectropolarimetrically in
the optical (e.g., the broad emission-line region or the
shielding gas itself could be responsible for the 
scattered light seen in the optical; \citealt{Ogle1999,Schmidt1999,Young2007}).
The scattering medium must be highly ionized so that there is no
strong Fe K$\alpha$ line produced in the scattering process
(see more discussion about ionization state in point 2 below).
In the disk-wind model of BAL quasars, the Compton-thin scattering medium could be
the hot low-density outflow in the polar region \citep[e.g.,][]{Proga2004},
although 
the column density and covering factor of the outflow
shown in the simulation results of \citet{Proga2004}
are probably insufficient to produce the $\approx6\%$ scattering 
fraction for diluting the line EW to the $\approx100$~eV level, and the 
simulations were not designed for radio-loud objects
with jets (such as PG~1004+130) either.
With such continuum dilution, the column density estimated
in Section~\ref{sec-columnd} is again likely a lower limit.

\item
The strength and centroid energy of the Fe K$\alpha$ line depend on the
ionization state of the scattering/reflecting medium
\citep[e.g.,][]{Matt1993,Ross1996,Kallman2004,Ross2005,Garcia2011}.
As the ionization parameter
($\xi$) increases, the EW of the line generally decreases and
the centroid energy increases; for a highly ionized medium
($\log\xi\approx4$), the EW of the Fe K$\alpha$ line can drop below 100~eV
\citep[e.g.,][]{Garcia2011}. In the disk-wind model of BAL quasars,
the scattering/reflecting medium is the shielding gas at the base of the
disk wind, and it is likely ionized and could be highly ionized
with $\log\xi\approx2$--6 \citep[e.g.,][]{Proga2004}.\footnote{Given 
the ionizing luminosity estimated from the intrinsic SED
(Figure~\ref{fig-sed}a), the location of the absorber ($\approx10^{16}$~cm),
the estimated column density ($\approx7\times10^{24}$~cm$^{-2}$),
and an assumed size of the absorber ($\approx10^{16}$~cm), we
estimated the ionization parameter to be $\log\xi\approx5$; this is likely
an overestimate, as we neglected the effects of Compton scattering in the
estimate and also the absorber could be located further out.}
Therefore,
a BAL quasar may be Compton-thick but without a strong Fe~K$\alpha$ line due to
highly ionized Compton-thick absorption, although it would appear difficult 
to ionize highly a Compton-thick medium due to the effects 
of Compton scattering (e.g., see Section 3 of \citealt{Schurch2009}).


\item
The strong narrow Fe K$\alpha$ line could be affected by line
broadening (or smeared out
in the extreme case) if the absorber/reflector has outflow motion and consists of
multiple velocity components.
While the kinematics of the \hbox{X-ray} absorbing component of the 
disk wind are still
uncertain, broadening and shifting of the Fe K$\alpha$ emission 
line has perhaps been seen in BAL quasars 
\citep[e.g.,][]{Oshima2001,Chartas2007,Sim2012}.

\end{enumerate}

\noindent We consider these explanations less likely than jet-linked dilution 
but still possible.

Of course, it is also possible that PG~1004+130 is not Compton-thick 
and thus does not have a prominent Fe K$\alpha$ line feature.
In this case, it would not produce \hbox{X-ray} emission
as typical quasars do and would be intrinsically \hbox{X-ray} weak, as discussed in
the next subsection.

\subsection{Intrinsic \hbox{X-ray} Weakness?}  \label{sec-xweak}

\subsubsection{Intrinsic \hbox{X-ray} Weakness in BAL Quasars
and Physical Implications}
PG~1004+130 and PG~1700+518 could also be intrinsically \hbox{X-ray} weak.
A small fraction of quasars have been suggested to be intrinsically \hbox{X-ray} weak
\citep[e.g.,][]{Gallagher2001,Sabra2001,Leighly2007,Wu2011,Miniutti2012}.
One of the best-studied cases is the \hbox{$z=0.192$}
narrow-line type 1 quasar PHL 1811, which is rapidly \hbox{X-ray} variable by a
high amplitude but always appears \hbox{X-ray} weak and shows no evidence
for intrinsic \hbox{X-ray} absorption \citep{Leighly2007}.
A systematic survey for such intrinsically \hbox{X-ray} weak
quasars demonstrated that these objects are rare in optically
selected (non-BAL) quasar samples; the fraction of SDSS quasars with
$\Delta\alpha_{\rm OX}<-0.4$
is $\la2\%$ \citep{Gibson2008}.

Many BAL quasars are \hbox{X-ray} weak due to absorption (see
Section 1.1), and thus they are often excluded in searches for
intrinsically \hbox{X-ray} weak quasars \citep[e.g.,][]{Gibson2008,Wu2011}. However,
intrinsic \hbox{X-ray} weakness is in fact an attractive possibility to account
for the BAL nature in the disk-wind scenario. As introduced in Section~1.1,
to launch successfully a wind through the UV line-driving mechanism,
the soft \hbox{X-ray} emission from the nucleus must be shielded to
prevent the wind from being overionized. If the nucleus were incapable of
producing strong \hbox{X-ray} emission, the wind could be launched
with little/no shielding, and then
a BAL quasar would be observed if the viewing angle were appropriate.
It is clear that some BAL quasars do emit \hbox{X-rays} at a nominal level
as their \hbox{X-ray} fluxes recover to expected levels after absorption corrections (see Section 1.1).
However, there are other BAL quasars,
including PG~1004+130 and PG~1700+518, that still appear \hbox{X-ray} weak
after basic absorption corrections; these are candidates for being
intrinsically \hbox{X-ray} weak
quasars.

The nature of intrinsic \hbox{X-ray} weakness remains unclear. 
Possible explanations generally invoke mechanisms that weaken or
destroy the \hbox{X-ray} emitting accretion-disk 
corona (see, e.g., the discussions in \citealt{Leighly2007} and 
\citealt{Miniutti2012}). 
One specific mechanism relevant to
BAL quasars was proposed by \citet{Proga2005}, who
suggested that part of the accretion-disk outflow (i.e., a dense, highly
ionized ``failed wind'' produced by overionization)
could fall into (see Figure~1 for the velocity field)
the corona and thereby suppress its \hbox{X-ray} emission.
With the presence of dense gas from the outflow,
the coronal magnetic field becomes insufficient to liberate and transport energy
from the disk to heat the corona via magnetized bubbles,
and also
relativistic electrons in the corona will cool efficiently via bremsstrahlung 
instead of inverse-Compton radiation
(bremsstrahlung is less effective in making hard X-rays).
We caution that the natures of both the ``failed wind'' and the corona
remain uncertain, and therefore there are inevitable uncertainties 
associated with the interaction of these two components

Considering the modeled dynamical nature 
of the outflow \citep[e.g.,][]{Proga2000} and the 
above coronal-quenching model, 
we suggest that there may even be a cyclical mechanism that 
switches on/off the coronal \hbox{X-ray} emission:
after the quenching of the coronal \hbox{X-ray} emission,
the wind can be successfully launched and there will be no ``failed wind'' 
falling into
the corona; the corona thus recovers to a standard \hbox{X-ray} emitting mode, and
overionizes the inner portion of wind which will then again fall into the corona
and suppress its \hbox{X-ray} emission.
The outflow is expected to settle down
to a steady state over a timescale of years
\citep[e.g.,][]{Proga2000}, and thus we might expect the above mechanism to 
operate
over a timescale of years or longer.
Such a cyclical mechanism could be used to explain why some BAL quasars
are \hbox{X-ray} normal (after absorption corrections) and some are perhaps
intrinsically \hbox{X-ray} weak, and it may be responsible for the 
significant \hbox{X-ray} flux and spectral variability observed in some BAL 
quasars \citep[e.g., PG~2112+059;][]{Gallagher2004}.
If BAL winds exist in all/most quasars, such a mechanism
could also be responsible for the strong \hbox{X-ray} variability
seen from some non-BAL quasars when they have entered \hbox{X-ray} weak states
due to quenched coronal emission
\citep[e.g., PHL 1092;][]{Miniutti2012}.

PG~1004+130 and PG~1700+518 appear to have some intrinsic \hbox{X-ray} absorption
regardless of whether they are intrinsically \hbox{X-ray} weak.
For PG~1004+130, the column density derived from spectral modeling is 
$N_{\rm H}=(1.8\pm0.6)\times10^{22}$~cm$^{-2}$ (see Section~\ref{sec-spec}).
For PG~1700+518, the {\it XMM-Newton} data reveal a column 
density of a few $10^{23}$~cm$^{-2}$ \citep{Ballo2011}; 
significant absorption is also indicated by the small 
effective photon index ($\Gamma_{\rm eff}\approx0.5\pm0.7$)
derived using the \nustar\ band ratio (see Section~\ref{sec-pho}).
Within the above coronal-quenching mechanism,
the apparent absorption could still be attributed to the 
shielding gas in BAL quasars as there may be a period when the 
coronal \hbox{X-ray} emission is quenched while the shielding gas has not fully
disappeared.
The values of $\Delta\alpha_{\rm OX}$ are both
$\approx-0.4$ for these two BAL quasars after corrections for apparent absorption
\citep{Miller2006,Ballo2011}, indicating that they are $\approx10$ times
intrinsically \hbox{X-ray} weaker than expected.
Intrinsic \hbox{X-ray} weakness would naturally explain the lack of a strong Fe 
K$\alpha$ line in PG~1004+130, as the observed 
X-ray spectra are not scattering/reflection dominated.
For the radio-loud object, PG~1004+130, 
the intrinsic \hbox{X-ray} weakness scenario implies that the radio jet
does not produce \hbox{X-ray} emission at a nominal level either ($\ga10$ times
weaker). 
Although unusual, a small fraction of radio-loud quasars appear to be providing
little/no excess \hbox{X-ray} emission compared to their radio-quiet counterparts
\citep[see, e.g., Figure 7 of][]{Miller2011}. The likely large inclination
angle of PG~1004+130 ($\ga45\degr$; e.g., \citealt{Wills1999,Miller2006}) 
could be one factor that reduces the observed X-ray emission compared
to other radio-loud quasars due to a lack of relativistic beaming.

\subsubsection{Intrinsic \hbox{X-ray} Weakness and Emission-Line Properties}
Intrinsic \hbox{X-ray} weakness in quasars may affect the appearance of the optical--UV 
emission lines by modifying the photoionization properties of 
the emission-line region. PHL~1811, for example, 
has unusual line emission:
its high-ionization lines (e.g., 
\ion{C}{4}~$\lambda1549$ and \ion{Si}{4}~$\lambda1400$) are very weak,
there is no evidence for forbidden or semiforbidden lines 
(e.g., [\ion{O}{3}]~$\lambda5007$), and the \ion{Fe}{2} and 
\ion{Fe}{3} pseudo-continuum in the near-UV is very strong \citep{Leighly2007b}.
The PHL 1811 analogs investigated by \citet{Wu2011} have similar emission-line
properties and are also \hbox{X-ray} weak quasar candidates.
Such unusual emission lines can probably be attributed to a soft  
ionizing continuum (UV nominal but \hbox{X-ray} weak; e.g., 
\citealt{Leighly2007b}).
For PG~1004+130 and PG~1700+518, the \ion{C}{4}~$\lambda1549$ and
\ion{Si}{4}~$\lambda1400$ lines are strongly absorbed due to
their BAL nature \citep[e.g.,][]{Wills1999,Brandt2000}, 
rendering reliable line-strength measurements impossible.
PG~1004+130 shows normal 
[\ion{O}{3}]~$\lambda5007$ emission; PG~1700+518 is a 
narrow-line type 1 quasar (as for PHL 1811) and shows no 
visible [\ion{O}{3}]~$\lambda5007$ line \citep[e.g.,][]{Boroson1992}. 
Their optical emission-line properties in general do not appear unusual compared 
to other PG quasars when considering ``Eigenvector 1'' correlations 
\citep[e.g.,][]{Boroson1992,Brandt1998}.

It is difficult to determine if
PG~1004+130 and PG~1700+518 share the same usual emission-line properties
as PHL 1811 when the weakness of their 
high-ionization lines cannot be constrained.
However, even if they have generally normal line emission, 
the intrinsic \hbox{X-ray} weakness scenario may still work, based on the following 
consideration. For a typical quasar,  
the ionizing continuum is dominated by EUV emission, which is
largely unobservable and is often estimated via interpolating the UV and 
\hbox{X-ray} photometric data points assuming a power-law spectrum 
(e.g., see the composite quasar SEDs in
Figure~\ref{fig-sed}; also adopted in \citealt{Leighly2007b}). 
The coronal-quenching mechanism discussed above could perhaps mainly
reduce the $\ga0.5$~keV \hbox{X-ray} emission while leaving the EUV radiation from the 
accretion disk largely the same. 
Although the significantly weakened \hbox{X-ray} emission would 
certainly affect the ionization state of the emission-line region 
somewhat, it is possible that the appearance of the emission 
lines is affected
less dramatically due to ionization from EUV photons; further photoionization
calculations of emission lines are required to assess this in detail.

\subsubsection{Statistical Constraints on the Fraction of Intrinsically 
\hbox{X-ray} Weak BAL Quasars}
\begin{figure*}
\centerline{
\includegraphics[scale=0.5]{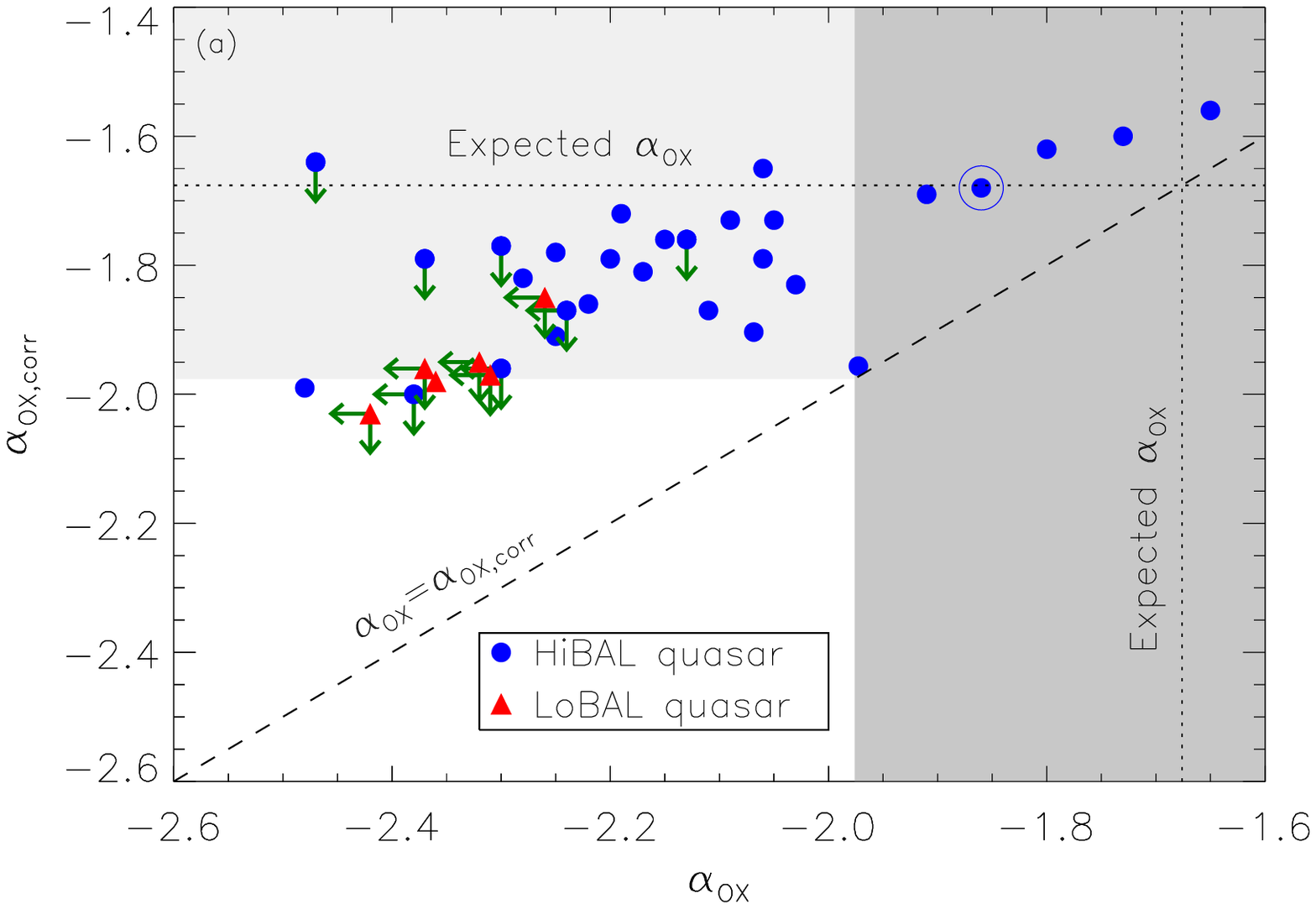}
\includegraphics[scale=0.5]{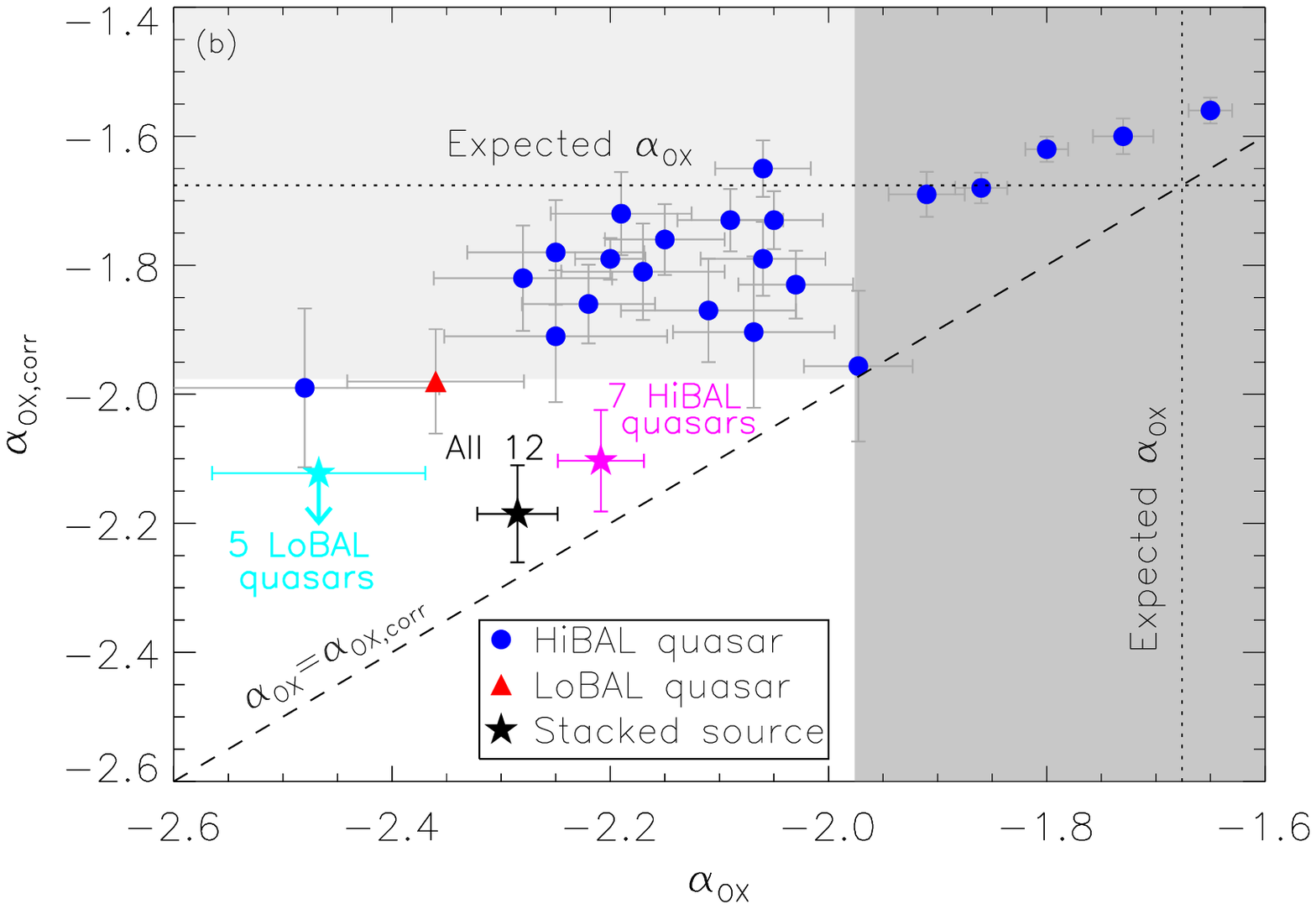}
}
\caption{
(a): $\alpha_{\rm OX}$ vs. $\alpha_{\rm OX,corr}$ for the 35 BAL quasars
in the \citet{Gallagher2006} sample.
HiBAL (29 objects) and LoBAL (6 objects) quasars are
shown as blue and red data points, respectively.
The open circle marks the only radio-loud quasar in this sample.
Upper limits on
$\alpha_{\rm OX}$ ($\alpha_{\rm OX,corr}$) are shown (green arrows) if there
are no detections in the soft (hard) X-ray band; the 
uncertainties on the other data points
are not shown in this panel for display purposes. The expected
$\alpha_{\rm OX}$ value was calculated using the
\citet{Steffen2006}
\hbox{$\alpha_{\rm OX}$--$L_{\rm 2500~\AA}$} relation and the
mean $L_{\rm 2500~\AA}$ for this sample. The dark and light shaded
regions show the 90\% confidence-level uncertainty associated with the expected $\alpha_{\rm OX}$.
(b): The same as panel (a), with the 12 upper-limit sources
replaced by their stacked source (black star).
The stacked source for the seven 
HiBAL (five LoBAL) quasars among the 12 sources 
is shown as the magenta star (cyan star).
\label{fig-aox06}}
\end{figure*}

Motivated by our \nustar\ results, we derived basic statistical constraints 
upon the fraction of intrinsically \hbox{X-ray} weak
BAL quasars from the relatively well-studied Large Bright
Quasar Survey (LBQS) BAL quasar
sample in \citet{Gallagher2006}. 
These are distant BAL quasars at $z\approx1.5$--3; at their mean
redshift of $z\approx2$, the 0.5--8~keV \chandra\ observations are probing the 
rest-frame $\approx1.5$--24~keV band, close
to the hard \hbox{X-ray} band observed by \nustar.
These quasars are all radio quiet except for one, 
and thus we do not expect significant 
jet-linked contributions to the X-ray emission 
as for the case of PG~1004+130.
We searched for intrinsically X-ray weak quasar candidates in these
sample sources
by comparing their $\alpha_{\rm OX}$ and $\alpha_{\rm OX,corr}$
(the absorption-corrected
$\alpha_{\rm OX}$; derived by assuming a
$\Gamma=2$ power-law spectrum
and normalizing it to the rest-frame \hbox{$\ga5$~keV} continuum) values from \citet{Gallagher2006}\footnote{We reinvestigated the source detections in 
the \citet{Gallagher2006} sample by 
running {\sc wavdetect} at low significance levels and then examining
the source significances with the 
binomial probability approach (see Equation 1). Two quasars 
not detected in the 2--8 keV band in \citet{Gallagher2006} were detected, and 
we computed their $\alpha_{\rm OX,corr}$ values instead of using upper limits.}
to the expected $\alpha_{\rm OX}$ values derived from the \citet{Steffen2006}
\hbox{$\alpha_{\rm OX}$--$L_{\rm 2500~\AA}$} relation.
These quasars have a limited range of $L_{\rm 2500~\AA}$ (with
standard deviation $\sigma=0.25$~dex), and thus we adopted the 
mean expected $\alpha_{\rm OX}$ value ($-1.68$) and its 
90\% confidence-level uncertainty 
($0.3$; from Figure~5 of \citealt{Gibson2008}) in the comparison.
We also performed the comparison by deriving $\Delta\alpha_{\rm OX}$ and 
$\Delta\alpha_{\rm OX,corr}$ for each object individually, and we got 
consistent results with those below.

Figure~\ref{fig-aox06}a shows the
$\alpha_{\rm OX}$ vs. $\alpha_{\rm OX,corr}$ plot for the 35 BAL
quasars in the sample. 
The shaded regions indicate the 
expected mean $\alpha_{\rm OX}$ for this sample 
and its 90\% uncertainty.
Sources that lie near the slanted dashed line ($\Gamma=2$ soft spectra,
while heavily obscured or Compton-thick spectra are usually hard) 
and outside the shaded regions (being X-ray weak)
are candidates for being intrinsically \hbox{X-ray} weak
quasars. 
A significant fraction of the sources
lie above the
$\alpha_{\rm OX}=\alpha_{\rm OX,corr}$ line, indicating that
X-ray absorption is likely present.
Only six objects have $\alpha_{\rm OX}$ values in the expected range
for typical quasars
(dark-shaded region). However, after absorption corrections,
15 additional objects have $\alpha_{\rm OX,corr}$ values
within the expected range (light-shaded region), indicating that
they probably have nominal underlying X-ray emission. There are four
objects below the light-shaded region (two measurements and two upper limits),
and another ten in
the light-shaded region have upper limits on $\alpha_{\rm OX,corr}$.
We consider these 14 objects as possible candidates
for being intrinsically \hbox{X-ray} weak
quasars, and thus the fraction is $\la40\%$ (i.e., $14/35$) among BAL quasars.

Such a constraint is relatively loose as 12 of the 14 candidates have
only upper limits on $\alpha_{\rm OX,corr}$. 
We performed a \chandra\ stacking analysis to obtain the average X-ray
properties of these 12 sources \citep[cf. Section 4.2 of][]{Wu2011}.
We added the total counts and background counts of the sources
extracted from $\approx95\%$ EEF apertures and larger annular 
background regions, respectively, in the observed 
0.5--2~keV (soft) and 2--8~keV (hard) bands.
For each band, we computed the binomial no-source probability (see
Equation 1) to determine if the stacked source is detected.
Using the Portable, Interactive, Multi-Mission Simulator
(PIMMS),\footnote{http://cxc.harvard.edu/toolkit/pimms.jsp.}
we derived the effective photon index from the ratio between
the soft-band and hard-band counts.
We adopted the average values of redshift, Galactic absorption
column density, and $f_{2500~{\rm \AA}}$ (there was only a 
small spread in $f_{2500~{\rm \AA}}$, 
with standard deviation $\sigma=0.24$~dex) to compute the 
$\alpha_{\rm OX}$ and $\alpha_{\rm OX,corr}$ parameters 
($\alpha_{\rm OX,corr}$ was derived using the 
hard-band flux and assuming $\Gamma=2$) for this subsample.

The stacking results are listed in Table~3, and the 
stacked data point
is shown as the black star in Figure~\ref{fig-aox06}b.
The stacked source is detected in both the soft and hard bands,
with net counts of $21.7^{+5.9}_{-4.8}$ and $5.6^{+3.8}_{-2.5}$ after
aperture corrections, respectively, and it has 
$\Gamma_{\rm eff}=1.6^{+0.6}_{-0.5}$.
Given the high luminosity ($>10^{43}$~\lum) in the soft band, the
observed soft X-ray spectrum (covering rest-frame $\approx1.5$--6~keV) 
should have negligible contribution from any
host-galaxy emission.
It appears that the stacked source is a candidate for 
being intrinsically X-ray weak, as it is close to the slanted dashed line
in the unshaded region (having a fairly soft X-ray spectrum and being X-ray weak).
As a first-order estimate, we consider that about 
half (six) of the 12 quasars being
stacked
are probably softer ($\Gamma_{\rm eff}\ga1.6$) than the stacked signal,
and the other half are harder 
($\Gamma_{\rm eff}\la1.6$).
All these objects are outside the dark-shaded region (X-ray
weak based on $\alpha_{\rm OX}$), while the six having a soft spectrum 
($\Gamma_{\rm eff}\ga1.6$) would 
lie close to the $\alpha_{\rm OX}$=$\alpha_{\rm OX,corr}$ line
if they could be detected individually.
Therefore, we expect that at least six ($\ga17\%$) of the 35 BAL 
quasars are candidates 
for being intrinsically \hbox{X-ray} weak.
Combined with the $\approx40\%$ upper limit,
we constrain the fraction to be $\approx17$--40\%.
Deeper \chandra\ observations of these 12 objects that detect them
individually could significantly narrow the estimated 
range of this fraction.

Among the 12 stacked objects, there are seven HiBAL and five LoBAL quasars
(see Footnote~\ref{footnote-bal}). As LoBAL quasars are X-ray weaker than
HiBAL quasars in general, 
we stacked these two groups separately, and the results are also shown 
in Table~3 and Figure~\ref{fig-aox06}b. The stacked signal of the
12 objects is dominated by the seven HiBAL quasars,
but both of these groups of seven HiBAL quasars 
and five LoBAL quasars could contain intrinsically X-ray weak
quasars, given their stacked X-ray properties and the positions of the stacked
sources in the $\alpha_{\rm OX}$ vs. $\alpha_{\rm OX,corr}$ plot.

The fraction of intrinsically X-ray weak objects among BAL quasars
estimated here ($\approx17$--40\%) is considerably larger 
than the $\la2\%$ fraction of 
intrinsically \hbox{X-ray} weak objects among non-BAL quasars
\citep{Gibson2008}.
If non-BAL quasars also have BAL winds lying out of the line of sight
(see Section~1.1),
the significant difference in the fractions would suggest that 
intrinsically \hbox{X-ray} weak quasars are preferentially seen as BAL
quasars. One possible scenario is that although the average
covering factor of the BAL wind is $\approx0.2$ (see Section~1.1),
the wind in an intrinsically \hbox{X-ray} weak quasar has
a considerably larger
covering factor, as it is likely easier to launch the wind when the nuclear
X-ray emission is weak. Therefore, we would tend to 
observe BALs preferentially 
in the spectra of intrinsically \hbox{X-ray} weak quasars. 

Compared to the SDSS BAL quasar sample, the LBQS BAL quasars
have somewhat higher optical luminosities (by $\approx0.5$~dex on average) 
and stronger BAL features, and they are preferentially X-ray weaker (e.g., see Section 4.6.2
of \citealt{Gibson2009}). The  
weaker X-ray emission and stronger BAL features
of LBQS BAL quasars could be related to
more absorption/coronal-quenching and thus stronger radiatively driven winds.
As the nature
of intrinsic X-ray weakness is highly uncertain, 
we caution that the constraints on the fraction ($\approx17$--40\%)
derived from the LBQS sample might not be applicable to the SDSS or
other BAL quasar
samples, and further X-ray studies of large BAL quasar samples are required 
to constrain the fraction better.

\section{SUMMARY, IMPLICATIONS, AND FUTURE WORK}

\subsection{Summary and Implications}
In this paper, we have investigated the hard \hbox{X-ray} emission observed by \nustar\ 
from two optically bright BAL quasars, PG~1004+130 and PG~1700+518,
and we have discussed the nature of 
their weak hard \hbox{X-ray} emission.
The key points from this work are summarized below:

\begin{enumerate}
\item
PG~1004+130 and PG~1700+518 have been observed by \nustar\ 
with exposure times
of $32.4$~ks and $82.5$~ks, respectively. 
PG~1004+130 was detected in both FPMs A and B, while
PG~1700+518 was only detected in FPM~A.
The \hbox{X-ray} positions of both objects are in good agreement with
their optical positions ($\la3\arcsec$), indicating 
the detections are reliable. See Section~\ref{sec-pho}.

\item
We provided aperture-photometry results
for the two targets in five \nustar\ bands:
\hbox{4--10~keV}, \hbox{4--20~keV}, \hbox{10--20~keV}, \hbox{20--30~keV},
and \hbox{30--79~keV}. The \hbox{X-ray} luminosities in the 4--20~keV 
band are $(5.3\pm0.8)\times10^{43}$~\lum\ for PG~1004+130
and $(2.3\pm0.6)\times10^{43}$~\lum\ for PG~1700+518.
We also derived an effective power-law photon index
based on the \nustar\ band ratio. PG~1004+130 is soft with
$\Gamma_{\rm eff}\approx1.7$, and PG~1700+518 is hard with
$\Gamma_{\rm eff}\approx0.5$. See Section~\ref{sec-pho}.

\item
We jointly analyzed the \nustar\ and \chandra\ spectra 
for PG~1004+130. The spectra were modeled with a partial-covering
absorber model. The resulting absorption is moderate,
$N_{\rm H}=(1.8\pm0.6)\times10^{22}$~cm$^{-2}$, and it is insufficient to 
explain the \hbox{X-ray} weakness of this quasar. Furthermore, it is 
likely that the 3--8~keV flux has decreased by a factor of $\approx2.3$
in the 2012 \nustar\ observation compared to its \chandra\ flux in 2005.
There is no Fe~K$\alpha$ emission line detected at rest-frame 6.4~keV,
with an upper limit on the rest-frame EW of $\approx178$~eV.
See Section~\ref{sec-spec}.

\item
We constructed radio-to-X-ray SEDs for PG~1004+130 and PG~1700+518.
From the radio to the UV,
the continuum SEDs of the two targets agree with
composite quasar SEDs (aside from dust reddening 
in PG~1700+518). Besides the significant \hbox{X-ray} weakness around rest-frame
2~keV, these two objects 
also appear more than an order of magnitude less luminous than typical quasars
at rest-frame 20~keV, and even the 40~keV luminosity upper limits
are below expectations. This hard \hbox{X-ray} weakness implies that the two 
BAL quasars either have Compton-thick absorption or are intrinsically \hbox{X-ray} weak.
See Section~\ref{sec-sed}.

\item
We derived column-density constraints using the {\sc MYTorus} model, 
under the assumption that 
the weak hard \hbox{X-ray} emission is caused by absorption.
For both objects, Compton-thick absorption appears required for any typical
assumption about the intrinsic $\alpha_{\rm OX}$ value.
The derived column densities are \hbox{$N_{\rm H}\approx7\times10^{24}$~cm$^{-2}$} for both 
BAL quasars, corresponding to Thomson optical depths of $\tau_{\rm T}\approx5$. 
Compton-thick absorption in the disk-wind model
requires the shielding gas to be located within $\approx10^{16}$~cm of the SMBH 
(assuming a standard accretion disk)
to prevent blocking of the UV radiation that drives the outflowing
wind.
We discussed jet-linked dilution and a few other 
possibilities that could cause the lack of a strong Fe~K$\alpha$ line
in PG~1004+130. See Section~\ref{sec-ct}.

\item
We discussed an intrinsic \hbox{X-ray} weakness scenario
that may relax the \hbox{X-ray} shielding requirement
for launching the 
accretion-disk wind in BAL quasars.
PG~1004+130 and PG~1700+518 are $\approx10$ times
intrinsically \hbox{X-ray} weaker than typical quasars under this scenario.
Based on the \citet{Proga2005} coronal-quenching model, we 
suggested a cyclical mechanism that could switch on/off the coronal \hbox{X-ray} 
emission and be responsible for a mix of 
intrinsically \hbox{X-ray} normal and
intrinsically \hbox{X-ray} weak BAL quasars. 
Motivated by our \nustar\ results, we
estimated the fraction of intrinsically \hbox{X-ray} weak
BAL quasars to be $\approx17$--40\% via 
a \chandra\ stacking analysis with the LBQS
BAL quasar sample.
See Section~\ref{sec-xweak}.

\end{enumerate}

Irrespective of its cause, the hard \hbox{X-ray} weakness of these two BAL 
quasars discovered by \nustar\
has implications for the detection and characterization 
of AGNs in deep \hbox{X-ray} surveys. 
PG~1004+130 and PG~1700+518 are among 
the most optically luminous BAL quasars known at low redshift (Figure~\ref{fig-lz}), 
and in each object substantial SMBH growth is clearly occuring.
However, we expect that PG~1700+518 could not be detected in a
600~ks \nustar\ survey\footnote{The 
deepest \nustar\ survey currently proposed, the 
\nustar\ Extended \chandra\ Deep Field-South survey, will have an
exposure of $\approx400$--800~ks.} 
if placed at {$z\ga0.7$}. In the deepest \chandra\
survey to date, the 4 Ms \chandra\ Deep Field-South \citep[e.g.,][]{Xue2011},
PG~1700+518 would be detectable if placed at high redshifts (e.g., $z\ga3$),
as \chandra\ is actually probing the hard \hbox{X-ray} bands at these redshifts.
However, with $\la100$ counts detected, the \chandra\ data could not constrain
its intrinsic spectrum accurately, and the derived \hbox{X-ray} properties 
would have a large uncertainty (e.g., the derived \hbox{X-ray} luminosity
and the amount of SMBH growth would be 
an order of magnitude lower than the real values).
Therefore, detection completeness and AGN characterization 
in deep \hbox{X-ray} surveys should be interpreted
carefully, considering the likely existence of X-ray weak BAL and related
quasars in the local and 
distant
universe \citep[e.g.,][]{Alexander2008,Burlon2011}.
In this case, a UV-excess selection of bright quasars that 
are X-ray weak could be utilized to search for and study the distant
counterparts of PG~1004+130 and PG~1700+518.

\subsection{Future Work}
Given the limited data for the two targets studied here,
we cannot strongly prefer the 
Compton-thick absorption scenario or the intrinsic \hbox{X-ray} weakness scenario.
A \nustar\ survey of a large sample 
of BAL quasars showing similar \hbox{X-ray} weakness 
(i.e., being significantly \hbox{X-ray} weak 
even after corrections for apparent absorption)
may help to discriminate 
between these two scenarios.
If heavy absorption is responsible for the \hbox{X-ray} weakness,
a continuous distribution of column densities would be expected, probably
ranging from $\ga5\times10^{23}$~cm$^{-2}$ to being significantly Compton-thick
($\approx10^{25}$~cm$^{-2}$). If 
all the observed survey targets show
weak hard \hbox{X-ray} emission like PG~1004+130 and PG~1700+518, 
this would indicate an unlikely scenario where
we had found an isolated population of highly Compton-thick objects via
soft \hbox{X-ray} selection, and thus we would consider that intrinsic \hbox{X-ray} 
weakness is probably the correct resolution. 
Alternatively, obtaining a hard \hbox{X-ray} spectrum for a similar
object with sufficient counts that
allows more detailed
spectral analysis may also provide some useful constraints that would
shed light on the nature of these BAL quasars.

We hypothesized a cyclical mechanism in Section~4.2.1
that could explain why some BAL quasars are intrinsically \hbox{X-ray} normal 
and others are perhaps intrinsically \hbox{X-ray} weak.
Further numerical simulations that carefully treat the temporal evolution 
of the \hbox{X-ray} 
shielding gas will be required to assess this dynamical 
model. An \hbox{X-ray} variability study of a large sample of BAL quasars over a long
timescale (years)
may also help to assess whether BAL quasars vary between \hbox{X-ray} normal and 
X-ray weak states \citep[see, e.g.,][]{Saez2012}. 
This could be achieved by snapshot monitoring observations
of a sample of \hbox{X-ray} weak BAL quasars using \chandra, {\it XMM-Newton}, 
and/or \nustar\ which would provide multi-epoch and large-bandpass coverage.

We estimated the fraction of intrinsically
X-ray weak BAL quasars ($\approx17$--40\%) in Section~4.2.3. 
This constraint could likely be 
tightened significantly if we could obtain \chandra\ hard-band (2--8~keV; 
$\approx6$--24~keV in the rest frame) flux measurements of the 
12 sources with only relatively weak upper-limit information presently.
These objects likely contain strong candidates for
intrinsically X-ray weak quasars based on the stacking analysis.
They were previously
observed by \chandra\ with 5--7~ks exposures \citep{Gallagher2006}.
It could be a useful investment to obtain additional $\approx20$--30~ks 
\chandra\ observations that will improve the hard-band 
detection limit by a factor of $\approx5$ and detect a significant
fraction of these 12 sources individually.

~\\
~\\

We acknowledge support from the California Institute of
Technology (Caltech) \nustar\ subcontract 44A-1092750 (BL, WNB),
NASA ADP grant NNX10AC99G (BL, WNB), 
the Leverhulme trust and the Science Technology and Facilities Council 
(DMA),
Basal-CATA grant PFB-06/2007 and
CONICYT-Chile grants FONDECYT 1101024 and Anillo ACT1101 (FEB),
and CONICYT-Chile grant FONDECYT 3120198 (CS).
We thank M.~Young for help with the planning of this project
and K.~Forster for help with the \nustar\ data access, and
we thank M.~Balokovic, K.~Boydstun, T.~N.~Lu, B.~P.~Miller,
Jianfeng Wu, and T. Yaqoob for helpful discussions.
We thank the referee, S.~C.~Gallagher, for carefully
reviewing the manuscript and providing helpful comments.

This work was supported under NASA Contract No. NNG08FD60C, and made use 
of data from the \nustar\ mission, a project led by Caltech, 
managed by the Jet Propulsion Laboratory, 
and funded by the National Aeronautics and Space Administration. 
We thank the \nustar\ Operations, Software and Calibration teams for 
support with the execution and analysis of these observations.
This research has made use of NuSTARDAS 
jointly developed by the ASI Science 
Data Center (ASDC, Italy) and Caltech (USA).

~\\


\begin{thebibliography}{146}
\expandafter\ifx\csname natexlab\endcsname\relax\def\natexlab#1{#1}\fi

\bibitem[{{Abazajian} {et~al.}(2009){Abazajian}, {Adelman-McCarthy},
  {Ag{\"u}eros}, {et~al.}}]{Abazajian2009}
{Abazajian}, K.~N., {Adelman-McCarthy}, J.~K., {Ag{\"u}eros}, M.~A., {et~al.}
  2009, \apjs, 182, 543

\bibitem[{{Alexander} {et~al.}(2008){Alexander}, {Chary}, {Pope},
  {et~al.}}]{Alexander2008}
{Alexander}, D.~M., {Chary}, R.-R., {Pope}, A., {et~al.} 2008, \apj, 687, 835

\bibitem[{{Allen} {et~al.}(2011){Allen}, {Hewett}, {Maddox}, {Richards}, \&
  {Belokurov}}]{Allen2011}
{Allen}, J.~T., {Hewett}, P.~C., {Maddox}, N., {Richards}, G.~T., \&
  {Belokurov}, V. 2011, \mnras, 410, 860

\bibitem[{{Anders} \& {Grevesse}(1989)}]{Anders1989}
{Anders}, E., \& {Grevesse}, N. 1989, \gca, 53, 197

\bibitem[{{Arnaud}(1996)}]{Arnaud1996}
{Arnaud}, K.~A. 1996, in ASP Conf. Ser., Vol. 101, Astronomical Data Analysis
  Software and Systems V, ed. G.~H. {Jacoby} \& J.~{Barnes}, 17

\bibitem[{{Ballo} {et~al.}(2011){Ballo}, {Piconcelli}, {Vignali}, \&
  {Schartel}}]{Ballo2011}
{Ballo}, L., {Piconcelli}, E., {Vignali}, C., \& {Schartel}, N. 2011, \mnras,
  415, 2600

\bibitem[{{Balucinska-Church} \& {McCammon}(1992)}]{Balucinska1992}
{Balucinska-Church}, M., \& {McCammon}, D. 1992, \apj, 400, 699

\bibitem[{{Bassani} {et~al.}(1999){Bassani}, {Dadina}, {Maiolino},
  {et~al.}}]{Bassani1999}
{Bassani}, L., {Dadina}, M., {Maiolino}, R., {et~al.} 1999, \apjs, 121, 473

\bibitem[{{Becker} {et~al.}(2000){Becker}, {White}, {Gregg},
  {et~al.}}]{Becker2000}
{Becker}, R.~H., {White}, R.~L., {Gregg}, M.~D., {et~al.} 2000, \apj, 538, 72

\bibitem[{{Blackburne} {et~al.}(2011){Blackburne}, {Pooley}, {Rappaport}, \&
  {Schechter}}]{Blackburne2011}
{Blackburne}, J.~A., {Pooley}, D., {Rappaport}, S., \& {Schechter}, P.~L. 2011,
  \apj, 729, 34

\bibitem[{{Borguet} {et~al.}(2013){Borguet}, {Arav}, {Edmonds}, {Chamberlain},
  \& {Benn}}]{Borguet2012}
{Borguet}, B.~C.~J., {Arav}, N., {Edmonds}, D., {Chamberlain}, C., \& {Benn},
  C. 2013, \apj, 762, 49

\bibitem[{{Boroson} \& {Green}(1992)}]{Boroson1992}
{Boroson}, T.~A., \& {Green}, R.~F. 1992, \apjs, 80, 109

\bibitem[{{Braito} {et~al.}(2004){Braito}, {Della Ceca}, {Piconcelli},
  {et~al.}}]{Braito2004}
{Braito}, V., {Della Ceca}, R., {Piconcelli}, E., {et~al.} 2004, \aap, 420, 79

\bibitem[{{Brandt} \& {Boller}(1998)}]{Brandt1998}
{Brandt}, N., \& {Boller}, T. 1998, Astronomische Nachrichten, 319, 7

\bibitem[{{Brandt} {et~al.}(2000){Brandt}, {Laor}, \& {Wills}}]{Brandt2000}
{Brandt}, W.~N., {Laor}, A., \& {Wills}, B.~J. 2000, \apj, 528, 637

\bibitem[{{Broos} {et~al.}(2007){Broos}, {Feigelson}, {Townsley},
  {et~al.}}]{Broos2007}
{Broos}, P.~S., {Feigelson}, E.~D., {Townsley}, L.~K., {et~al.} 2007, \apjs,
  169, 353

\bibitem[{{Brotherton} {et~al.}(2001){Brotherton}, {Arav}, {Becker},
  {et~al.}}]{Brotherton2001}
{Brotherton}, M.~S., {Arav}, N., {Becker}, R.~H., {et~al.} 2001, \apj, 546, 134

\bibitem[{{Burlon} {et~al.}(2011){Burlon}, {Ajello}, {Greiner},
  {et~al.}}]{Burlon2011}
{Burlon}, D., {Ajello}, M., {Greiner}, J., {et~al.} 2011, \apj, 728, 58

\bibitem[{{Calzetti} {et~al.}(2000){Calzetti}, {Armus}, {Bohlin}, {Kinney},
  {Koornneef}, \& {Storchi-Bergmann}}]{Calzetti2000}
{Calzetti}, D., {Armus}, L., {Bohlin}, R.~C., {Kinney}, A.~L., {Koornneef}, J.,
  \& {Storchi-Bergmann}, T. 2000, \apj, 533, 682

\bibitem[{{Capellupo} {et~al.}(2011){Capellupo}, {Hamann}, {Shields},
  {Rodr{\'{\i}}guez Hidalgo}, \& {Barlow}}]{Capellupo2011}
{Capellupo}, D.~M., {Hamann}, F., {Shields}, J.~C., {Rodr{\'{\i}}guez Hidalgo},
  P., \& {Barlow}, T.~A. 2011, \mnras, 413, 908

\bibitem[{{Capellupo} {et~al.}(2012){Capellupo}, {Hamann}, {Shields},
  {Rodr{\'{\i}}guez Hidalgo}, \& {Barlow}}]{Capellupo2012}
---. 2012, \mnras, 422, 3249

\bibitem[{{Chartas} {et~al.}(2007){Chartas}, {Eracleous}, {Dai}, {Agol}, \&
  {Gallagher}}]{Chartas2007}
{Chartas}, G., {Eracleous}, M., {Dai}, X., {Agol}, E., \& {Gallagher}, S. 2007,
  \apj, 661, 678

\bibitem[{{Chartas} {et~al.}(2009){Chartas}, {Saez}, {Brandt}, {Giustini}, \&
  {Garmire}}]{Chartas2009}
{Chartas}, G., {Saez}, C., {Brandt}, W.~N., {Giustini}, M., \& {Garmire}, G.~P.
  2009, \apj, 706, 644

\bibitem[{{Clavel} {et~al.}(2006){Clavel}, {Schartel}, \& {Tomas}}]{Clavel2006}
{Clavel}, J., {Schartel}, N., \& {Tomas}, L. 2006, \aap, 446, 439

\bibitem[{{Comastri}(2004)}]{Comastri2004}
{Comastri}, A. 2004, in Astrophysics and Space Science Library, Vol. 308,
  Supermassive Black Holes in the Distant Universe, ed. A.~J. {Barger}, 245

\bibitem[{{Condon} {et~al.}(1998){Condon}, {Cotton}, {Greisen},
  {et~al.}}]{Condon1998}
{Condon}, J.~J., {Cotton}, W.~D., {Greisen}, E.~W., {et~al.} 1998, \aj, 115,
  1693

\bibitem[{{Crenshaw} {et~al.}(1999){Crenshaw}, {Kraemer}, {Boggess}, {Maran},
  {Mushotzky}, \& {Wu}}]{Crenshaw1999}
{Crenshaw}, D.~M., {Kraemer}, S.~B., {Boggess}, A., {Maran}, S.~P.,
  {Mushotzky}, R.~F., \& {Wu}, C.-C. 1999, \apj, 516, 750

\bibitem[{{Di Matteo} {et~al.}(2005){Di Matteo}, {Springel}, \&
  {Hernquist}}]{Dimatteo2005}
{Di Matteo}, T., {Springel}, V., \& {Hernquist}, L. 2005, \nat, 433, 604

\bibitem[{{Dickey} \& {Lockman}(1990)}]{Dickey1990}
{Dickey}, J.~M., \& {Lockman}, F.~J. 1990, \araa, 28, 215

\bibitem[{{DiPompeo} {et~al.}(2013){DiPompeo}, {Brotherton}, \& {De
  Breuck}}]{DiPompeo2012}
{DiPompeo}, M.~A., {Brotherton}, M.~S., \& {De Breuck}, C. 2013, \mnras, 428,
  1565

\bibitem[{{Elvis} \& {Fabbiano}(1984)}]{Elvis1984}
{Elvis}, M., \& {Fabbiano}, G. 1984, \apj, 280, 91

\bibitem[{{Elvis} {et~al.}(2012){Elvis}, {Hao}, {Civano}, {et~al.}}]{Elvis2012}
{Elvis}, M., {Hao}, H., {Civano}, F., {et~al.} 2012, \apj, 759, 6

\bibitem[{{Elvis} {et~al.}(1994){Elvis}, {Wilkes}, {McDowell},
  {et~al.}}]{Elvis1994}
{Elvis}, M., {Wilkes}, B.~J., {McDowell}, J.~C., {et~al.} 1994, \apjs, 95, 1

\bibitem[{{Emmering} {et~al.}(1992){Emmering}, {Blandford}, \&
  {Shlosman}}]{Emmering1992}
{Emmering}, R.~T., {Blandford}, R.~D., \& {Shlosman}, I. 1992, \apj, 385, 460

\bibitem[{{Eracleous} {et~al.}(2000){Eracleous}, {Sambruna}, \&
  {Mushotzky}}]{Eracleous2000}
{Eracleous}, M., {Sambruna}, R., \& {Mushotzky}, R.~F. 2000, \apj, 537, 654

\bibitem[{{Evans} \& {Koratkar}(2004)}]{Evans2004}
{Evans}, I.~N., \& {Koratkar}, A.~P. 2004, \apjs, 150, 73

\bibitem[{{Fan} {et~al.}(2009){Fan}, {Wang}, {Wang}, {et~al.}}]{Fan2009}
{Fan}, L.~L., {Wang}, H.~Y., {Wang}, T., {et~al.} 2009, \apj, 690, 1006

\bibitem[{{Filiz Ak} {et~al.}(2012){Filiz Ak}, {Brandt}, {Hall},
  {et~al.}}]{Filiz2012}
{Filiz Ak}, N., {Brandt}, W.~N., {Hall}, P.~B., {et~al.} 2012, \apj, 757, 114

\bibitem[{{Freeman} {et~al.}(2002){Freeman}, {Kashyap}, {Rosner}, \&
  {Lamb}}]{Freeman2002}
{Freeman}, P.~E., {Kashyap}, V., {Rosner}, R., \& {Lamb}, D.~Q. 2002, \apjs,
  138, 185

\bibitem[{{Gallagher} {et~al.}(2002){Gallagher}, {Brandt}, {Chartas}, \&
  {Garmire}}]{Gallagher2002a}
{Gallagher}, S.~C., {Brandt}, W.~N., {Chartas}, G., \& {Garmire}, G.~P. 2002,
  \apj, 567, 37

\bibitem[{{Gallagher} {et~al.}(2006){Gallagher}, {Brandt}, {Chartas},
  {et~al.}}]{Gallagher2006}
{Gallagher}, S.~C., {Brandt}, W.~N., {Chartas}, G., {et~al.} 2006, \apj, 644,
  709

\bibitem[{{Gallagher} {et~al.}(2001){Gallagher}, {Brandt}, {Laor},
  {et~al.}}]{Gallagher2001}
{Gallagher}, S.~C., {Brandt}, W.~N., {Laor}, A., {et~al.} 2001, \apj, 546, 795

\bibitem[{{Gallagher} {et~al.}(1999){Gallagher}, {Brandt}, {Sambruna},
  {Mathur}, \& {Yamasaki}}]{Gallagher1999}
{Gallagher}, S.~C., {Brandt}, W.~N., {Sambruna}, R.~M., {Mathur}, S., \&
  {Yamasaki}, N. 1999, \apj, 519, 549

\bibitem[{{Gallagher} {et~al.}(2004){Gallagher}, {Brandt}, {Wills},
  {et~al.}}]{Gallagher2004}
{Gallagher}, S.~C., {Brandt}, W.~N., {Wills}, B.~J., {et~al.} 2004, \apj, 603,
  425

\bibitem[{{Gallagher} {et~al.}(2007){Gallagher}, {Hines}, {Blaylock},
  {et~al.}}]{Gallagher2007}
{Gallagher}, S.~C., {Hines}, D.~C., {Blaylock}, M., {et~al.} 2007, \apj, 665,
  157

\bibitem[{{Ganguly} \& {Brotherton}(2008)}]{Ganguly2008}
{Ganguly}, R., \& {Brotherton}, M.~S. 2008, \apj, 672, 102

\bibitem[{{Garc{\'{\i}}a} {et~al.}(2011){Garc{\'{\i}}a}, {Kallman}, \&
  {Mushotzky}}]{Garcia2011}
{Garc{\'{\i}}a}, J., {Kallman}, T.~R., \& {Mushotzky}, R.~F. 2011, \apj, 731,
  131

\bibitem[{{Garmire} {et~al.}(2003){Garmire}, {Bautz}, {Ford}, {Nousek}, \&
  {Ricker}}]{Garmire2003}
{Garmire}, G.~P., {Bautz}, M.~W., {Ford}, P.~G., {Nousek}, J.~A., \& {Ricker},
  Jr., G.~R. 2003, Proc. SPIE, 4851, 28

\bibitem[{{Gehrels}(1986)}]{Gehrels1986}
{Gehrels}, N. 1986, \apj, 303, 336

\bibitem[{{Ghisellini} {et~al.}(1994){Ghisellini}, {Haardt}, \&
  {Matt}}]{Ghisellini1994}
{Ghisellini}, G., {Haardt}, F., \& {Matt}, G. 1994, \mnras, 267, 743

\bibitem[{{Gibson} {et~al.}(2010){Gibson}, {Brandt}, {Gallagher}, {Hewett}, \&
  {Schneider}}]{Gibson2010}
{Gibson}, R.~R., {Brandt}, W.~N., {Gallagher}, S.~C., {Hewett}, P.~C., \&
  {Schneider}, D.~P. 2010, \apj, 713, 220

\bibitem[{{Gibson} {et~al.}(2008{\natexlab{a}}){Gibson}, {Brandt}, \&
  {Schneider}}]{Gibson2008}
{Gibson}, R.~R., {Brandt}, W.~N., \& {Schneider}, D.~P. 2008{\natexlab{a}},
  \apj, 685, 773

\bibitem[{{Gibson} {et~al.}(2008{\natexlab{b}}){Gibson}, {Brandt}, {Schneider},
  \& {Gallagher}}]{Gibson2008b}
{Gibson}, R.~R., {Brandt}, W.~N., {Schneider}, D.~P., \& {Gallagher}, S.~C.
  2008{\natexlab{b}}, \apj, 675, 985

\bibitem[{{Gibson} {et~al.}(2009){Gibson}, {Jiang}, {Brandt},
  {et~al.}}]{Gibson2009}
{Gibson}, R.~R., {Jiang}, L., {Brandt}, W.~N., {et~al.} 2009, \apj, 692, 758

\bibitem[{{Giustini} {et~al.}(2008){Giustini}, {Cappi}, \&
  {Vignali}}]{Giustini2008}
{Giustini}, M., {Cappi}, M., \& {Vignali}, C. 2008, \aap, 491, 425

\bibitem[{{Grandi} {et~al.}(2006){Grandi}, {Malaguti}, \&
  {Fiocchi}}]{Grandi2006}
{Grandi}, P., {Malaguti}, G., \& {Fiocchi}, M. 2006, \apj, 642, 113

\bibitem[{{Green} {et~al.}(2001){Green}, {Aldcroft}, {Mathur}, {Wilkes}, \&
  {Elvis}}]{Green2001}
{Green}, P.~J., {Aldcroft}, T.~L., {Mathur}, S., {Wilkes}, B.~J., \& {Elvis},
  M. 2001, \apj, 558, 109

\bibitem[{{Gregg} {et~al.}(2006){Gregg}, {Becker}, \& {de Vries}}]{Gregg2006}
{Gregg}, M.~D., {Becker}, R.~H., \& {de Vries}, W. 2006, \apj, 641, 210

\bibitem[{{Grupe} {et~al.}(2003){Grupe}, {Mathur}, \& {Elvis}}]{Grupe2003}
{Grupe}, D., {Mathur}, S., \& {Elvis}, M. 2003, \aj, 126, 1159

\bibitem[{{G{\"u}ltekin} {et~al.}(2009){G{\"u}ltekin}, {Richstone}, {Gebhardt},
  {et~al.}}]{Gultekin2009}
{G{\"u}ltekin}, K., {Richstone}, D.~O., {Gebhardt}, K., {et~al.} 2009, \apj,
  698, 198

\bibitem[{{Guyon} {et~al.}(2006){Guyon}, {Sanders}, \& {Stockton}}]{Guyon2006}
{Guyon}, O., {Sanders}, D.~B., \& {Stockton}, A. 2006, \apjs, 166, 89

\bibitem[{{Haas} {et~al.}(2003){Haas}, {Klaas}, {M{\"u}ller},
  {et~al.}}]{Haas2003}
{Haas}, M., {Klaas}, U., {M{\"u}ller}, S.~A.~H., {et~al.} 2003, \aap, 402, 87

\bibitem[{{Harrison} {et~al.}(2013){Harrison}, {Craig}, {Christensen},
  {et~al.}}]{Harrison2013}
{Harrison}, F.~A., {Craig}, W.~W., {Christensen}, F.~E., {et~al.} 2013, \apj,
  770, 103

\bibitem[{{Hewett} \& {Foltz}(2003)}]{Hewett2003}
{Hewett}, P.~C., \& {Foltz}, C.~B. 2003, \aj, 125, 1784

\bibitem[{{Ikeda} {et~al.}(2009){Ikeda}, {Awaki}, \& {Terashima}}]{Ikeda2009}
{Ikeda}, S., {Awaki}, H., \& {Terashima}, Y. 2009, \apj, 692, 608

\bibitem[{{Jim{\'e}nez-Vicente} {et~al.}(2012){Jim{\'e}nez-Vicente},
  {Mediavilla}, {Mu{\~n}oz}, \& {Kochanek}}]{Jimenez2012}
{Jim{\'e}nez-Vicente}, J., {Mediavilla}, E., {Mu{\~n}oz}, J.~A., \& {Kochanek},
  C.~S. 2012, \apj, 751, 106

\bibitem[{{Just} {et~al.}(2007){Just}, {Brandt}, {Shemmer},
  {et~al.}}]{Just2007}
{Just}, D.~W., {Brandt}, W.~N., {Shemmer}, O., {et~al.} 2007, \apj, 665, 1004

\bibitem[{{Kallman} {et~al.}(2004){Kallman}, {Palmeri}, {Bautista}, {Mendoza},
  \& {Krolik}}]{Kallman2004}
{Kallman}, T.~R., {Palmeri}, P., {Bautista}, M.~A., {Mendoza}, C., \& {Krolik},
  J.~H. 2004, \apjs, 155, 675

\bibitem[{{Komatsu} {et~al.}(2011){Komatsu}, {Smith}, {Dunkley},
  {et~al.}}]{Komatsu2011}
{Komatsu}, E., {Smith}, K.~M., {Dunkley}, J., {et~al.} 2011, \apjs, 192, 18

\bibitem[{{Konigl} \& {Kartje}(1994)}]{Konigl1994}
{Konigl}, A., \& {Kartje}, J.~F. 1994, \apj, 434, 446

\bibitem[{{Kraft} {et~al.}(1991){Kraft}, {Burrows}, \& {Nousek}}]{Kraft1991}
{Kraft}, R.~P., {Burrows}, D.~N., \& {Nousek}, J.~A. 1991, \apj, 374, 344

\bibitem[{{LaMassa} {et~al.}(2011){LaMassa}, {Heckman}, {Ptak},
  {et~al.}}]{LaMassa2011}
{LaMassa}, S.~M., {Heckman}, T.~M., {Ptak}, A., {et~al.} 2011, \apj, 729, 52

\bibitem[{{Laor} \& {Brandt}(2002)}]{Laor2002}
{Laor}, A., \& {Brandt}, W.~N. 2002, \apj, 569, 641

\bibitem[{{Laor} {et~al.}(1997){Laor}, {Fiore}, {Elvis}, {Wilkes}, \&
  {McDowell}}]{Laor1997}
{Laor}, A., {Fiore}, F., {Elvis}, M., {Wilkes}, B.~J., \& {McDowell}, J.~C.
  1997, \apj, 477, 93

\bibitem[{{Lazarova} {et~al.}(2012){Lazarova}, {Canalizo}, {Lacy}, \&
  {Sajina}}]{Lazarova2012}
{Lazarova}, M.~S., {Canalizo}, G., {Lacy}, M., \& {Sajina}, A. 2012, \apj, 755,
  29

\bibitem[{{Leighly} {et~al.}(2007{\natexlab{a}}){Leighly}, {Halpern},
  {Jenkins}, \& {Casebeer}}]{Leighly2007b}
{Leighly}, K.~M., {Halpern}, J.~P., {Jenkins}, E.~B., \& {Casebeer}, D.
  2007{\natexlab{a}}, \apjs, 173, 1

\bibitem[{{Leighly} {et~al.}(2007{\natexlab{b}}){Leighly}, {Halpern},
  {Jenkins}, {et~al.}}]{Leighly2007}
{Leighly}, K.~M., {Halpern}, J.~P., {Jenkins}, E.~B., {et~al.}
  2007{\natexlab{b}}, \apj, 663, 103

\bibitem[{{Luo} {et~al.}(2013){Luo}, {Fabbiano}, {Strader}, {et~al.}}]{Luo2013}
{Luo}, B., {Fabbiano}, G., {Strader}, J., {et~al.} 2013, \apjs, 204, 14

\bibitem[{{Lusso} {et~al.}(2010){Lusso}, {Comastri}, {Vignali},
  {et~al.}}]{Lusso2010}
{Lusso}, E., {Comastri}, A., {Vignali}, C., {et~al.} 2010, \aap, 512, A34

\bibitem[{{Lynds}(1967)}]{Lynds1967}
{Lynds}, C.~R. 1967, \apj, 147, 396

\bibitem[{{Lyons}(1991)}]{Lyons1991}
{Lyons}, L. 1991, Data Analysis for Physical Science Students (Cambridge:
  Cambridge Univ. Press)

\bibitem[{{Martin} {et~al.}(2005){Martin}, {Fanson}, {Schiminovich},
  {et~al.}}]{Martin2005}
{Martin}, D.~C., {Fanson}, J., {Schiminovich}, D., {et~al.} 2005, \apjl, 619,
  L1

\bibitem[{{Mathur} {et~al.}(2000){Mathur}, {Green}, {Arav},
  {et~al.}}]{Mathur2000}
{Mathur}, S., {Green}, P.~J., {Arav}, N., {et~al.} 2000, \apjl, 533, L79

\bibitem[{{Matt} {et~al.}(1996){Matt}, {Brandt}, \& {Fabian}}]{Matt1996}
{Matt}, G., {Brandt}, W.~N., \& {Fabian}, A.~C. 1996, \mnras, 280, 823

\bibitem[{{Matt} {et~al.}(1993){Matt}, {Fabian}, \& {Ross}}]{Matt1993}
{Matt}, G., {Fabian}, A.~C., \& {Ross}, R.~R. 1993, \mnras, 262, 179

\bibitem[{{Miller} {et~al.}(2006){Miller}, {Brandt}, {Gallagher},
  {et~al.}}]{Miller2006}
{Miller}, B.~P., {Brandt}, W.~N., {Gallagher}, S.~C., {et~al.} 2006, \apj, 652,
  163

\bibitem[{{Miller} {et~al.}(2009){Miller}, {Brandt}, {Gibson}, {Garmire}, \&
  {Shemmer}}]{Miller2009}
{Miller}, B.~P., {Brandt}, W.~N., {Gibson}, R.~R., {Garmire}, G.~P., \&
  {Shemmer}, O. 2009, \apj, 702, 911

\bibitem[{{Miller} {et~al.}(2011){Miller}, {Brandt}, {Schneider},
  {et~al.}}]{Miller2011}
{Miller}, B.~P., {Brandt}, W.~N., {Schneider}, D.~P., {et~al.} 2011, \apj, 726,
  20

\bibitem[{{Miniutti} {et~al.}(2012){Miniutti}, {Brandt}, {Schneider},
  {et~al.}}]{Miniutti2012}
{Miniutti}, G., {Brandt}, W.~N., {Schneider}, D.~P., {et~al.} 2012, \mnras,
  425, 1718

\bibitem[{{Murphy}(2009)}]{Murphy2009b}
{Murphy}, K. 2009, PhD thesis, The Johns Hopkins University

\bibitem[{{Murphy} \& {Yaqoob}(2009)}]{Murphy2009}
{Murphy}, K.~D., \& {Yaqoob}, T. 2009, \mnras, 397, 1549

\bibitem[{{Murray} {et~al.}(1995){Murray}, {Chiang}, {Grossman}, \&
  {Voit}}]{Murray1995}
{Murray}, N., {Chiang}, J., {Grossman}, S.~A., \& {Voit}, G.~M. 1995, \apj,
  451, 498

\bibitem[{{Neugebauer} {et~al.}(1987){Neugebauer}, {Green}, {Matthews},
  {et~al.}}]{Neugebauer1987}
{Neugebauer}, G., {Green}, R.~F., {Matthews}, K., {et~al.} 1987, \apjs, 63, 615

\bibitem[{{Neugebauer} {et~al.}(1979){Neugebauer}, {Oke}, {Becklin}, \&
  {Matthews}}]{Neugebauer1979}
{Neugebauer}, G., {Oke}, J.~B., {Becklin}, E.~E., \& {Matthews}, K. 1979, \apj,
  230, 79

\bibitem[{{Ogle} {et~al.}(1999){Ogle}, {Cohen}, {Miller}, {et~al.}}]{Ogle1999}
{Ogle}, P.~M., {Cohen}, M.~H., {Miller}, J.~S., {et~al.} 1999, \apjs, 125, 1

\bibitem[{{Oshima} {et~al.}(2001){Oshima}, {Mitsuda}, {Fujimoto},
  {et~al.}}]{Oshima2001}
{Oshima}, T., {Mitsuda}, K., {Fujimoto}, R., {et~al.} 2001, \apjl, 563, L103

\bibitem[{{Page} {et~al.}(2005){Page}, {Reeves}, {O'Brien}, \&
  {Turner}}]{Page2005}
{Page}, K.~L., {Reeves}, J.~N., {O'Brien}, P.~T., \& {Turner}, M.~J.~L. 2005,
  \mnras, 364, 195

\bibitem[{{Peterson} {et~al.}(2004){Peterson}, {Ferrarese}, {Gilbert},
  {et~al.}}]{Peterson2004}
{Peterson}, B.~M., {Ferrarese}, L., {Gilbert}, K.~M., {et~al.} 2004, \apj, 613,
  682

\bibitem[{{Pettini} \& {Boksenberg}(1985)}]{Pettini1985}
{Pettini}, M., \& {Boksenberg}, A. 1985, \apjl, 294, L73

\bibitem[{{Piconcelli} {et~al.}(2013){Piconcelli}, {Miniutti}, {Ranalli},
  {et~al.}}]{Piconcelli2013}
{Piconcelli}, E., {Miniutti}, G., {Ranalli}, P., {et~al.} 2013, \mnras, 428,
  1185

\bibitem[{{Pooley} {et~al.}(2007){Pooley}, {Blackburne}, {Rappaport}, \&
  {Schechter}}]{Pooley2007}
{Pooley}, D., {Blackburne}, J.~A., {Rappaport}, S., \& {Schechter}, P.~L. 2007,
  \apj, 661, 19

\bibitem[{{Proga}(2005)}]{Proga2005}
{Proga}, D. 2005, \apjl, 630, L9

\bibitem[{{Proga} \& {Kallman}(2004)}]{Proga2004}
{Proga}, D., \& {Kallman}, T.~R. 2004, \apj, 616, 688

\bibitem[{{Proga} {et~al.}(2000){Proga}, {Stone}, \& {Kallman}}]{Proga2000}
{Proga}, D., {Stone}, J.~M., \& {Kallman}, T.~R. 2000, \apj, 543, 686

\bibitem[{{Reeves} {et~al.}(1997){Reeves}, {Turner}, {Ohashi}, \&
  {Kii}}]{Reeves1997}
{Reeves}, J.~N., {Turner}, M.~J.~L., {Ohashi}, T., \& {Kii}, T. 1997, \mnras,
  292, 468

\bibitem[{{Reynolds}(1997)}]{Reynolds1997}
{Reynolds}, C.~S. 1997, \mnras, 286, 513

\bibitem[{{Richards} {et~al.}(2006){Richards}, {Lacy}, {Storrie-Lombardi},
  {et~al.}}]{Richards2006}
{Richards}, G.~T., {Lacy}, M., {Storrie-Lombardi}, L.~J., {et~al.} 2006, \apjs,
  166, 470

\bibitem[{{Ross} \& {Fabian}(2005)}]{Ross2005}
{Ross}, R.~R., \& {Fabian}, A.~C. 2005, \mnras, 358, 211

\bibitem[{{Ross} {et~al.}(1996){Ross}, {Fabian}, \& {Brandt}}]{Ross1996}
{Ross}, R.~R., {Fabian}, A.~C., \& {Brandt}, W.~N. 1996, \mnras, 278, 1082

\bibitem[{{Rupke} \& {Veilleux}(2013)}]{Rupke2013}
{Rupke}, D.~S.~N., \& {Veilleux}, S. 2013, \apj, 768, 75

\bibitem[{{Sabra} \& {Hamann}(2001)}]{Sabra2001}
{Sabra}, B.~M., \& {Hamann}, F. 2001, \apj, 563, 555

\bibitem[{{Saez} {et~al.}(2012){Saez}, {Brandt}, {Gallagher}, {Bauer}, \&
  {Garmire}}]{Saez2012}
{Saez}, C., {Brandt}, W.~N., {Gallagher}, S.~C., {Bauer}, F.~E., \& {Garmire},
  G.~P. 2012, \apj, 759, 42

\bibitem[{{Schartel} {et~al.}(2005){Schartel}, {Rodr{\'{\i}}guez-Pascual},
  {Santos-Lle{\'o}}, {et~al.}}]{Schartel2005}
{Schartel}, N., {Rodr{\'{\i}}guez-Pascual}, P.~M., {Santos-Lle{\'o}}, M.,
  {et~al.} 2005, \aap, 433, 455

\bibitem[{{Schmidt} \& {Hines}(1999)}]{Schmidt1999}
{Schmidt}, G.~D., \& {Hines}, D.~C. 1999, \apj, 512, 125

\bibitem[{{Schmidt} \& {Green}(1983)}]{Schmidt1983}
{Schmidt}, M., \& {Green}, R.~F. 1983, \apj, 269, 352

\bibitem[{{Schneider} {et~al.}(2010){Schneider}, {Richards}, {Hall},
  {et~al.}}]{Schneider2010}
{Schneider}, D.~P., {Richards}, G.~T., {Hall}, P.~B., {et~al.} 2010, \aj, 139,
  2360

\bibitem[{{Schurch} {et~al.}(2009){Schurch}, {Done}, \& {Proga}}]{Schurch2009}
{Schurch}, N.~J., {Done}, C., \& {Proga}, D. 2009, \apj, 694, 1

\bibitem[{{Scott} {et~al.}(2011){Scott}, {Stewart}, {Mateos},
  {et~al.}}]{Scott2011}
{Scott}, A.~E., {Stewart}, G.~C., {Mateos}, S., {et~al.} 2011, \mnras, 417, 992

\bibitem[{{Serjeant} \& {Hatziminaoglou}(2009)}]{Serjeant2009}
{Serjeant}, S., \& {Hatziminaoglou}, E. 2009, \mnras, 397, 265

\bibitem[{{Shakura} \& {Sunyaev}(1973)}]{Shakura1973}
{Shakura}, N.~I., \& {Sunyaev}, R.~A. 1973, \aap, 24, 337

\bibitem[{{Shang} {et~al.}(2011){Shang}, {Brotherton}, {Wills},
  {et~al.}}]{Shang2011}
{Shang}, Z., {Brotherton}, M.~S., {Wills}, B.~J., {et~al.} 2011, \apjs, 196, 2

\bibitem[{{Shankar} {et~al.}(2008){Shankar}, {Dai}, \&
  {Sivakoff}}]{Shankar2008}
{Shankar}, F., {Dai}, X., \& {Sivakoff}, G.~R. 2008, \apj, 687, 859

\bibitem[{{Shemmer} {et~al.}(2005){Shemmer}, {Brandt}, {Gallagher},
  {et~al.}}]{Shemmer2005}
{Shemmer}, O., {Brandt}, W.~N., {Gallagher}, S.~C., {et~al.} 2005, \aj, 130,
  2522

\bibitem[{{Shemmer} {et~al.}(2008){Shemmer}, {Brandt}, {Netzer}, {Maiolino}, \&
  {Kaspi}}]{Shemmer2008}
{Shemmer}, O., {Brandt}, W.~N., {Netzer}, H., {Maiolino}, R., \& {Kaspi}, S.
  2008, \apj, 682, 81

\bibitem[{{Sim} {et~al.}(2012){Sim}, {Proga}, {Kurosawa}, {et~al.}}]{Sim2012}
{Sim}, S.~A., {Proga}, D., {Kurosawa}, R., {et~al.} 2012, \mnras, 426, 2859

\bibitem[{{Sim} {et~al.}(2010){Sim}, {Proga}, {Miller}, {Long}, \&
  {Turner}}]{Sim2010}
{Sim}, S.~A., {Proga}, D., {Miller}, L., {Long}, K.~S., \& {Turner}, T.~J.
  2010, \mnras, 408, 1396

\bibitem[{{Skrutskie} {et~al.}(2006){Skrutskie}, {Cutri}, {Stiening},
  {et~al.}}]{Skrutskie2006}
{Skrutskie}, M.~F., {Cutri}, R.~M., {Stiening}, R., {et~al.} 2006, \aj, 131,
  1163

\bibitem[{{Sprayberry} \& {Foltz}(1992)}]{Sprayberry1992}
{Sprayberry}, D., \& {Foltz}, C.~B. 1992, \apj, 390, 39

\bibitem[{{Steffen} {et~al.}(2006){Steffen}, {Strateva}, {Brandt},
  {et~al.}}]{Steffen2006}
{Steffen}, A.~T., {Strateva}, I., {Brandt}, W.~N., {et~al.} 2006, \aj, 131,
  2826

\bibitem[{{Sturm} {et~al.}(2011){Sturm}, {Gonz{\'a}lez-Alfonso}, {Veilleux},
  {et~al.}}]{Sturm2011}
{Sturm}, E., {Gonz{\'a}lez-Alfonso}, E., {Veilleux}, S., {et~al.} 2011, \apjl,
  733, L16

\bibitem[{{Trump} {et~al.}(2006)}]{Trump2006}
{Trump}, J.~R., {et~al.} 2006, \apjs, 165, 1

\bibitem[{{Turner} {et~al.}(1997){Turner}, {George}, {Nandra}, \&
  {Mushotzky}}]{Turner1997}
{Turner}, T.~J., {George}, I.~M., {Nandra}, K., \& {Mushotzky}, R.~F. 1997,
  \apj, 488, 164

\bibitem[{{Vanden Berk} {et~al.}(2001){Vanden Berk}, {Richards}, {Bauer},
  {et~al.}}]{Vandenberk2001}
{Vanden Berk}, D.~E., {Richards}, G.~T., {Bauer}, A., {et~al.} 2001, \aj, 122,
  549

\bibitem[{{Vestergaard} \& {Peterson}(2006)}]{Vestergaard2006}
{Vestergaard}, M., \& {Peterson}, B.~M. 2006, \apj, 641, 689

\bibitem[{{Weymann} {et~al.}(1991){Weymann}, {Morris}, {Foltz}, \&
  {Hewett}}]{Weymann1991}
{Weymann}, R.~J., {Morris}, S.~L., {Foltz}, C.~B., \& {Hewett}, P.~C. 1991,
  \apj, 373, 23

\bibitem[{{White} \& {Becker}(1992)}]{White1992}
{White}, R.~L., \& {Becker}, R.~H. 1992, \apjs, 79, 331

\bibitem[{{Wills} {et~al.}(1999){Wills}, {Brandt}, \& {Laor}}]{Wills1999}
{Wills}, B.~J., {Brandt}, W.~N., \& {Laor}, A. 1999, \apjl, 520, L91

\bibitem[{{Wright} {et~al.}(2010){Wright}, {Eisenhardt}, {Mainzer},
  {et~al.}}]{Wright2010}
{Wright}, E.~L., {Eisenhardt}, P.~R.~M., {Mainzer}, A.~K., {et~al.} 2010, \aj,
  140, 1868

\bibitem[{{Wu} {et~al.}(2010){Wu}, {Brandt}, {Comins}, {et~al.}}]{Wu2010}
{Wu}, J., {Brandt}, W.~N., {Comins}, M.~L., {et~al.} 2010, \apj, 724, 762

\bibitem[{{Wu} {et~al.}(2011){Wu}, {Brandt}, {Hall}, {et~al.}}]{Wu2011}
{Wu}, J., {Brandt}, W.~N., {Hall}, P.~B., {et~al.} 2011, \apj, 736, 28

\bibitem[{{Xue} {et~al.}(2011){Xue}, {Luo}, {Brandt}, {et~al.}}]{Xue2011}
{Xue}, Y.~Q., {Luo}, B., {Brandt}, W.~N., {et~al.} 2011, \apjs, 195, 10

\bibitem[{{Yaqoob} \& {Murphy}(2009)}]{Yaqoob2009}
{Yaqoob}, T., \& {Murphy}, K. 2009, in Chandra's First Decade of Discovery, ed.
  S.~{Wolk}, A.~{Fruscione}, \& D.~{Swartz}, 57

\bibitem[{{Yaqoob} {et~al.}(2010){Yaqoob}, {Murphy}, {Miller}, \&
  {Turner}}]{Yaqoob2010}
{Yaqoob}, T., {Murphy}, K.~D., {Miller}, L., \& {Turner}, T.~J. 2010, \mnras,
  401, 411

\bibitem[{{York} {et~al.}(2000){York}, {Adelman}, {Anderson},
  {et~al.}}]{York2000}
{York}, D.~G., {Adelman}, J., {Anderson}, Jr., J.~E., {et~al.} 2000, \aj, 120,
  1579

\bibitem[{{Young} {et~al.}(2010){Young}, {Elvis}, \& {Risaliti}}]{Young2010}
{Young}, M., {Elvis}, M., \& {Risaliti}, G. 2010, \apj, 708, 1388

\bibitem[{{Young} {et~al.}(2007){Young}, {Axon}, {Robinson}, {Hough}, \&
  {Smith}}]{Young2007}
{Young}, S., {Axon}, D.~J., {Robinson}, A., {Hough}, J.~H., \& {Smith}, J.~E.
  2007, \nat, 450, 74

\end{thebibliography}

\begin{deluxetable}{lccccccc}
\tablecaption{\nustar\ Observation Log}

\tablehead{
\colhead{Object}                   &
\colhead{$z$}                   &
\colhead{Observation}                   &
\colhead{Observation}                   &
\colhead{Exp}                   &
\colhead{Exp\_clean}                   &
\colhead{$\Delta_{\rm OX}$}                &
\colhead{$N_{\rm H,Gal}$}     \\
\colhead{Name}                   &
\colhead{}                   &
\colhead{Start Date}                   &
\colhead{ID}                   &
\colhead{(ks)}                   &
\colhead{(ks)}                   &
\colhead{(arcsec)}  &
\colhead{$(10^{20}$~cm$^{-2})$}  \\
\colhead{(1)}         &
\colhead{(2)}         &
\colhead{(3)}         &
\colhead{(4)}         &
\colhead{(5)}         &
\colhead{(6)}         &
\colhead{(7)}         &
\colhead{(8)}
}

\startdata
PG~1004+130     &0.241&         2012 Oct 29&    60001112002 & 32.4&   30.1& 0.1 & 3.7 \\
PG~1700+518 &0.292&         2012 Sep 22&  60001113002&   82.5&   77.1&1.5 & 2.6 
\enddata

\tablecomments{
Cols. (1) and (2): object name and redshift.
Cols. (3) and (4): \nustar\ observation start date and observation ID.
Cols. (5) and (6): nominal and cleaned \nustar\ exposure times, respectively.
Col. (7): minimum positional offset between optical and \hbox{X-ray} positions.
The \hbox{X-ray} positions are determined using {\sc wavdetect} in the 4--20~keV 
images of FPMs A and B.
Col. (8): Galactic neutral hydrogen column density \citep{Dickey1990}.}
\label{tbl-obs}
\end{deluxetable}

\begin{deluxetable}{lccccccccccc}
\tablecaption{Photometric Properties}
\tablehead{
\colhead{Object Name}                   &
\multicolumn{5}{c}{Net Counts} &
\colhead{$\Gamma_{\rm eff}$\tablenotemark{a}}                   &
\multicolumn{4}{c}{Flux ($10^{-14}$~\flux)}             &
\colhead{$\log L$ (\lum)}   \\
\\\colhead{and FPM} & \cline{1-5} \cline{8-11} \\
\colhead{}                   &
\colhead{4--10}                   &
\colhead{4--20}                   &
\colhead{10--20}                   &
\colhead{20--30}                   &
\colhead{30--79}                   &
\colhead{}                  &
\colhead{4--10}                   &
\colhead{10--20}                   &
\colhead{20--30}                   &
\colhead{30--79}                  &
\colhead{4--20}  \\
\colhead{}  &
\colhead{keV}  &
\colhead{keV}  &
\colhead{keV}  &
\colhead{keV}  &
\colhead{keV}  &
\colhead{}  &
\colhead{keV}  &
\colhead{keV}  &
\colhead{keV}  &
\colhead{keV}  &
\colhead{keV}                   
}

\startdata
PG~1004+130 A    &$    112.1_{-14.2}^{+15.4}$&$    155.9_{-18.0}^{+19.2}$&$     44.7_{-11.0}^{+12.2}$&$ <14.4$&$ <25.7$&$        1.7\pm0.4$&$  15.4\pm2.0$&$  15.0\pm3.9$&$< 16.3$&$< 144.2$& $43.7\pm0.1$\\
PG~1004+130 B    &$    112.5_{-15.6}^{+16.8}$&$    153.2_{-19.3}^{+20.5}$&$     40.5_{-11.3}^{+12.5}$&$ <25.4$&$ <23.8$&$        1.8_{-0.4}^{+0.5}$&$  15.6\pm2.3$&$  13.4\pm4.0$&$< 28.6$&$< 129.8$& $43.7\pm0.1$\\
PG~1700+518 A & $     45.7_{-13.6}^{+15.8}$&$     86.7_{-18.4}^{+20.6}$&$     42.9_{-12.5}^{+14.7}$&$ <13.4$&$ <63.5$&$        0.5\pm0.7$&$   2.4\pm0.8$&$   6.0\pm1.9$&$<  6.0$&$< 172.4$&$43.3\pm0.1$\\
PG~1700+518 B &$                    <58.7$&$                    <89.4$&$                    <43.6$&$ <21.6$&$ <24.2$&$        0.5$&$<   3.1$&$  < 6.1$&$<  9.7$&$<  65.2$&$<43.4$
\enddata
\tablenotetext{a}{The effective photon index ($\Gamma_{\rm eff}$) was
derived based on the band ratio between
the observed 10--20~keV and 4--10~keV counts, assuming a power-law model with
Galactic absorption. See Section~\ref{sec-pho} for details. For PG~1700+518
in FPM~B, the value of $\Gamma_{\rm eff}$ was adopted as the one in 
FPM~A.}
\label{tbl-pho}
\end{deluxetable}

\begin{deluxetable}{cccccccccc}
\tablecaption{Stacked X-ray Properties for the Hard-Band Undetected LBQS BAL Quasars}
\tablehead{
\colhead{Sources} &
\colhead{Number of} &
\colhead{Mean} &
\colhead{Total Stacked} &
\colhead{Soft-Band} &
\colhead{Hard-Band} &
\colhead{$\Gamma_{\rm eff}$}                   &
\colhead{$\log L_{0.5\textrm{--}2\textrm{keV,rest}}$}  &
\colhead{$\alpha_{\rm OX}$} &
\colhead{$\alpha_{\rm OX,corr}$} \\
\colhead{Stacked} &
\colhead{Sources} &
\colhead{Redshift} &
\colhead{Exposure (ks)} &
\colhead{Counts} &
\colhead{Counts} &
\colhead{}                   &
\colhead{(\lum)}   &
\colhead{}                   &
\colhead{}                   
}
\startdata
All    & 12&1.99 & 68.4 &$21.7^{+5.9}_{-4.8}$ & $5.6^{+3.8}_{-2.5}$  &$1.6_{-0.5}^{+0.6}$& 43.3 & $-2.29\pm0.04$ &$-2.19\pm0.08$\\
HiBAL quasars    & 7&2.10 & 40.2 &  $18.7^{+5.6}_{-4.4}$ & $4.8^{+3.6}_{-2.3}$  &$1.6_{-0.5}^{+0.6}$& 43.5 & $-2.21\pm0.04$ &$-2.10\pm0.08$\\
LoBAL quasars    & 5&1.84 & 28.2 &  $3.0^{+3.1}_{-1.7}$ & $<3.8$\tablenotemark{a} &$>0.28$\tablenotemark{b}& 42.9 & $-2.47\pm0.10$ &$<-2.12$
\enddata
\tablenotetext{a}{The upper limit on the source counts was derived
using the Bayesian approach of \citet{Kraft1991} for a 90\% confidence level.}
\tablenotetext{b}{We assumed $\Gamma_{\rm eff}=2.0$ when calculating 
the soft-band
flux and $\alpha_{\rm OX}$.}
\label{tbl-stack}
\end{deluxetable}

\end{document}